\documentclass[
notitlepage,
bibnotes,
amsmath,amssymb,
aps,
10pt,
twocolumn,
prl
]{revtex4-2}
\usepackage{amssymb,bm,bbold,bbm,amsmath,mathtools}
\usepackage{xcolor}
\colorlet{mylinkcolor}{blue!66!black!80}

\newcommand{\e}{{\rm e}}

\newcommand{\Sigmaline}{\overline{\Sigma}_t}

\newcommand{\ps}{p_{\textrm{s}}}
\newcommand{\Var}{{\mathrm{Var}}}
\newcommand{\tmin}{t_{\textrm{min}}}
\newcommand{\nmin}{n_{\textrm{min}}}
\newcommand{\vs}{v_{\textrm{s}}}
\newcommand{\js}{j_{\textrm{s}}}

\DeclareMathOperator{\E}{\mathbb{E}}
\newcommand{\Sigmaloc}{\Sigma_\mathrm{loc}}

\usepackage{amsthm}

\usepackage{mathtools}
\usepackage[colorlinks=true,linkcolor=mylinkcolor,citecolor=mylinkcolor,filecolor=cyan,urlcolor=mylinkcolor,breaklinks=true]{hyperref}
\usepackage[utf8]{inputenc}
\usepackage{mlmodern}
\usepackage{orcidlink}
\usepackage{layouts}
\usepackage{microtype}

\begin{document}
\title{Thermodynamic Concentration Inequalities:\\  
Controlling Uncertainty in Finite-Time and Small-Sample Thermodynamic Inference}
\author{Rick Bebon\,\orcidlink{0000-0003-2187-0008}}
\email{rick.bebon@physik.uni-freiburg.de}
\author{Alja\v{z} Godec\,\orcidlink{0000-0003-1888-6666}}
\email{agodec@physik.uni-freiburg.de}
\affiliation{Mathematical Physics and Stochastic Dynamics, Faculty of Mathematics and Physics, University of Freiburg}

\begin{abstract}
We derive nonasymptotic upper bounds on the probability that a generalized current 
of a geometrically ergodic diffusion observed for any amount of time, or its sample mean over any arbitrary sample size, deviates from the stationary mean 
by more than any given amount. The concentration-of-measure behavior of generalized currents 
is universally governed by the relaxation time of the underlying dynamics, the locally observed dissipation rate, and the intrinsic local fluctuations of the observable. We uncover stark \emph{qualitative and quantitative} differences in fluctuations in and out of thermodynamic equilibrium. We further obtain \emph{refined} inverse thermodynamic uncertainty relations, bounding the variance of generalized currents from above. We construct nonasymptotic confidence intervals for controlling uncertainty in thermodynamic inference from small data, i.e., from short trajectories and small samples, and provide the first \emph{quantitative answer} to when a trajectory is sufficiently long and a sample is sufficiently large. As an illustration, we apply our results to currents observed on a two-dimensional Ornstein-Uhlenbeck process in and out of equilibrium and show how they can be used to rigorously detect broken detailed balance.
\end{abstract}
\maketitle

Additive functionals of stochastic sample paths---often referred to as
time-averaged observables~\cite{Burov2011PCCP,lapolla2018unfolding,lapolla2020spectral,dieball2022mathematical,dieball2022coarse,stutzer2026stochastic}---are
central to the analysis of stochastic time series~\cite{Time_series_analysis,Rivera_2018,Climate}, where time-averaging
along a
single or a handful of sufficiently long trajectories serves as a proxy for the
traditional ensemble average.
On the theoretical side, 
\emph{Stratonovich-type} functionals
$\overline{J}_t\equiv t^{-1}\int_0^t U(X_s)\circ d X_s$, so-called
generalized currents, have moved to the center of nonequilibrium statistical
mechanics 
and stochastic thermodynamics, where they define
energy-like quantities 
of mesoscopic systems along individual trajectories~\cite{sekimoto2010,seifert2012stochastic,seifert2025stochastic,peliti2021stochastic},
and extend the continuity equation
for additive path functionals~\cite{dieball2022coarse,dieball2022mathematical}.
Finite-time fluctuations of $\overline{J}_t$ obey
fluctuation theorems~\cite{seifert2012stochastic,Jarzynski2011,Evans2002}, and their tails encode rare events 
of thermodynamic significance, e.g., apparent ``violations'' of the second law~\cite{Wang2002,Jarzynski2011}.
Fluctuations are further constrained by
the thermodynamic
uncertainty relation (TUR) 
\cite{barato2015,gingrich2016dissipation,horowitz2020thermodynamic,dieball2023direct} 
which imposes a noise floor on any
current estimate.

Fluctuations of 
$\overline{J}_t$
are therefore 
intrinsically interesting and 
crucial for
interpreting single-molecule and particle-tracking
measurements~\cite{rief1997,ritort2006,rief2002force,hughes2016,xie1996single,ambrose1999single,woodside2014reconstructing,neuman2008single,camunas2016elastic,saxton2008,metzler2014,ernst2014,shen2017single}, as well as for controlling uncertainty 
in thermodynamic inference~\cite{seifert2019,horowitz2020thermodynamic,Dieball_2025_P,dieball2023direct}.
Specifically, TUR-based 
inference hinges on 
sampling the second and
higher-order
moments~\cite{dechant2020,Dechant2021,Wampler2021,Manikandan2022,Ray2023}
of the current functional, and 
inference of free-energy
differences from 
nonequilibrium work 
measurements via the Jarzynski~\cite{jarzynski1997nonequilibrium}
or Crooks~\cite{crooks1999} relations requires sufficient sampling of
work distribution \emph{tails}~\cite{collin2005verification,alemany2015free,ribezzi2014free,Pohorille2010,Hummer2001,Zuckerman2002,Jarzynski2006,Gore2003,engel2009asymptotics}.

\begin{figure}[!h]
    \centering
    \includegraphics{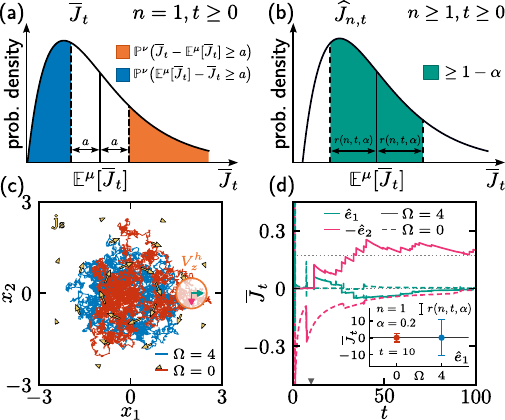}
    \caption{
    (a) Schematic probability density of $\overline{J}_t$
    for a single trajectory of finite length $t\geq0$.
    Fluctuations around the mean $\E^\mu[\overline{J}_t]$
    are quantified by right (orange) and left (blue)
    tail probabilities.
    (b) Uncertainty quantification. 
    The sample mean $\widehat{J}_{n,t}$ over $n$ trajectories of length $t$
    deviates from $\E^\mu[\overline{J}_t]$ by less than the confidence radius
    $r(n,t,\alpha)$ with probability at least $1-\alpha$ (green).
    (c) Trajectories of length $t=7.5$ for a driven ($\Omega=4$, blue) and equilibrium ($\Omega=0$, red)
    Ornstein-Uhlenbeck process with identical $\ps$ and $\lambda_\mathrm{gap}$.
    Yellow arrows depict $\js$, and $V^h_z$ (orange) probes coarse-grained currents
    along $\hat e_1$ (green) and $-\hat e_2$ (magenta).
    (d) Corresponding currents $\overline{J}_t$ 
    approach their stationary means 
    (dotted) only slowly.
    Inset: Error bars of a single $\overline{J}_t$ along $\hat e_1$ at $t=1$ 
    showing stark differences in (red) and out of (blue) equilibrium.
    }
    \label{fig:fig1}
\end{figure}

Reliable inference of 
the stationary mean $\E^\mu[\overline{J}_t]$ from a
time average or, more generally, from the 
sample mean
$\widehat{J}_{n,t}\equiv n^{-1}\sum_{i=1}^{n}\overline{J}^{(i)}_t$ over $n$
statistically independent paths of length $t$ 
hinges on a trajectory being ``sufficiently long'' and a sample
``sufficiently large''. What is ``sufficient'' is far from trivial \emph{a priori} 
and subsampling poses an additional challenge \cite{bebon2023controlling}.
Central-limit 
and bootstrap methods
may underestimate the
uncertainty and 
fail to guarantee 
coverage of the confidence
level~\cite{schenker1985qualms,Davison1997, Shao2003,hogg1995introduction},
while Bayesian approaches 
(contrary to common
perception~\cite{gelman1995bayesian,Lista2017,mcelreath2020statistical,kaplan2014bayesian}) do \emph{not} evade the small-sample problem,
as they remain sensitive to, and 
biased by, the 
prior choice~\cite{Smid2019}.
A systematic understanding of the 
fluctuations of $\overline{J}_t$
and 
$\widehat{J}_{n,t}$
remains elusive. 

Required but missing 
is nonasymptotic control of the \emph{full tails} of
$\overline{J}_t$ at finite $t$ and of $\widehat{J}_{n,t}$ at finite $n$. 
Indeed, the
large deviation description of current
fluctuations~\cite{maes2008steady,barato2015formal,chetrite2015nonequilibrium,touchette2009large,touchette2018introduction}
is 
asymptotic in $t$ and requires solving a nontrivial
spectral problem for a non-self-adjoint tilted
generator~\cite{chetrite2015nonequilibrium,touchette2018introduction}, which 
is
tractable
only for simple 
models~\cite{mehl2008large,fischer2018large,angeletti2016diffusions}
and otherwise necessitates 
numerical or sampling methods~\cite{giardina2006direct,lecomte2007numerical,ferre2018adaptive}.
Exact finite-time results 
are limited to variances and
correlations~\cite{dieball2022mathematical,dieball2022coarse}, and the TUR
and its inverse (iTUR)~\cite{bakewell2023general,vo2025inverse} constrain 
only 
the second moment.
Concentration inequalities 
provide nonasymptotic tail control 
but existing 
literature is limited to density-type functionals 
\cite{wu2000deviation,lezaud2001chernoff,cattiaux2008deviation,guillin2009transportation,gao2014bernstein,Birrell2025}
or, for currents, to Markov jump processes~\cite{bakewell2023general}.
A concentration analysis of Stratonovich-type functionals of continuous state-space
dynamics remains, to 
our knowledge, unexplored.

The 
intriguing question is whether (and how)
physical properties of the underlying dynamics $X_t$---in particular the entropy production---bound tail probabilities of $\overline{J}_t$ and $\widehat{J}_{n,t}$ from 
\emph{above}, complementary to the 
TUR limiting 
fluctuations 
from below.
Closely related is 
whether the fluctuations of
$\overline{J}_t$ differ \emph{qualitatively} between 
systems in and 
out of equilibrium (see Fig.~\ref{fig:fig1}d). 
Whereas the 
mean 
$\E^\mu[\overline{J}_t]$
vanishes under detailed balance, 
fluctuations persist,
and it is unclear 
how
irreversibility 
imprints on the statistics 
of $\overline{J}_t$
beyond the mean. 
Notwithstanding recent progress \cite{Roldn2010,Gladrow2016,Battle2016,Berezhkovskii2020,Gnesotto2018}, 
finite-time and small-sample fluctuations
make a
rigorous quantification of 
broken detailed balance
a daunting task. 

Here we derive nonasymptotic upper bounds---so-called concentration
inequalities~\cite{boucheron2013concentration}---on the probability that any
generalized current (including the mean 
entropy production rate) inferred from a trajectory of length $t$ of a geometrically ergodic
(and generally irreversible) diffusion process $X_t$ deviates from the
stationary mean $\E^\mu[\overline{J}_t]$ by more than $a$ (see Fig.~\ref{fig:fig1}a),
\begin{align}
\begin{rcases}
  \mathbb{P}^\nu\!\bigl( \overline{J}_t-\E^\mu[\overline{J}_t]\ge a\bigr)\\
  \mathbb{P}^\nu\!\bigl(\E^\mu[\overline{J}_t]- \overline{J}_t\ge a\bigr)
  \end{rcases}
  \leq N_\nu \exp\bigl[-t\mathcal{I}^J(a)\bigr],
  \label{eq:conc_intro}
\end{align}
valid for all $t>0$, $a\ge 0$, 
where 
$N_\nu$ is a prefactor accounting for 
arbitrary initial conditions $\nu$~\footnote{We assume $\nu$ to be absolutely continuous with respect to the invariant measure $\mu$ and $d\nu/d\mu \in L^2(\mu)$.}.
The 
function $\mathcal{I}^J(a)$
encodes the dynamics only through physically meaningful
quantities---the locally observed mean entropy production rate $\Sigma^U$, the spectral gap
$\lambda_{\mathrm{gap}}$, and the observable $U$ through 
its intrinsic local fluctuation $Q\equiv U\cdot D U$---which motivates naming Eq.~\eqref{eq:conc_intro} a \emph{thermodynamic concentration inequality}~\footnote{Note that compared to \cite{Hasegawa2024Thermodynamic}, we here derive upper bounds that control deviations in terms of genuinely thermodynamic quantities.}.
Applied to the entropy production, 
our bound shows that the mean dissipation 
bounds its fluctuations via a \emph{dissipation concentration inequality} [see Eq.~\eqref{eq:diss_conc_ineq}].
These bounds extend to the 
sample mean $\widehat{J}_{n,t}$, i.e., 
$\mathbb{P}^\nu(\widehat{J}_{n,t}-\E^\mu[\overline{J}_t]\ge a)
\leq N_\nu^n \exp[-nt\mathcal{I}^J(a)]$, for all $n\ge 1$ and $t>0$,
and allow 
to quantify the uncertainty via a confidence radius $r(n,t,\alpha)$, such that
$\E^\mu[\overline{J}_t]\in[\widehat{J}_{n,t}-r,\widehat{J}_{n,t}+r]$ with
probability of at least $1-\alpha$ (see Fig.~\ref{fig:fig1}b).
We obtain explicit results on the
minimal observation time $\tmin$ and sample size $\nmin$,
guaranteeing $r\leq\varepsilon$ for all $t\geq \tmin$ and $n\geq \nmin$, for a prescribed accuracy $\varepsilon$ and confidence
level $1-\alpha$. 
As a byproduct, our concentration-of-measure viewpoint
yields \emph{inverse thermodynamic uncertainty
relations}~\cite{bakewell2023general, vo2025inverse}---upper bounds on current variances 
complementary to the
TUR bounding them from below---and 
refines 
the recent result for overdamped Langevin
dynamics~\cite{vo2025inverse}.

Throughout, we demonstrate the validity and sharpness of our bounds
on the
two-dimensional Ornstein-Uhlenbeck process (OUP) (Fig.~\ref{fig:fig1}c,d) with irreversible
driving $\Omega$ that leaves the invariant density 
and spectral gap unchanged (see SM \cite{SM} for details).

\emph{Setup.---}We consider geometrically ergodic time-homogeneous 
dynamics $(X_t)_{t\geq 0}$ on the continuous state space
$\mathbb{R}^d$ (or a bounded domain 
with periodic or reflecting boundary conditions), 
governed by the stochastic
differential (overdamped Langevin) equation
\begin{align}
    d X_t = b(X_t) d t + \sigma d W_t,
\end{align}
with  (Lipschitz) drift 
$b:\mathbb{R}^d\to\mathbb{R}^d$, constant 
noise amplitude
$\sigma\in\mathbb{R}^{d\times m}$, and $(W_t)_{t\ge0}$ a standard $m$-dimensional
Wiener process.
The noise 
is related to the positive-definite diffusion matrix via $D = \sigma \sigma^\top/2$.
The corresponding probability density $p(x,t)$, 
with initial condition $p(x,0)$, 
evolves according to the Fokker-Planck equation
\begin{align}
    \partial_t p(x,t) = \mathcal{L}^\ast p(x,t)= -\nabla\cdot j(x,t),
\end{align}
where $j(x,t)= [b - D\nabla]p(x,t)$ is the probability current and $\mathcal{L}^\ast$ the $L^2(d x)$ adjoint of
the 
generator $\mathcal{L}$~\cite{gardiner1985handbook, pavliotis2014stochastic}
\begin{align}
    \mathcal{L}f &= b(x) \cdot \nabla f + \nabla \cdot (D\nabla f).
    \label{eq:markov_generator}
\end{align}
Ergodicity ensures relaxation to a unique steady state with invariant
density $\ps(x)\equiv\lim_{t\to\infty}p(x,t)$ and divergence-free
steady-state current $\js(x)\equiv[b(x)-D\nabla]\ps(x)$, defining the local
mean velocity $\vs(x)\equiv\js(x)/\ps(x)$. The steady state is 
out of equilibrium
whenever $\js\not\equiv0$ and obeys
detailed balance otherwise. 
Throughout, 
we write
$d \mu=\ps(x)d x$ for the invariant measure, and $\E^\mu$,
$\Var_\mu$, and $\langle\cdot,\cdot\rangle_\mu$ for the corresponding
expectation, variance, and $L^2(\mu)$ inner product.

The generator $\mathcal{L}=\mathcal{L}_A + \mathcal{L}_S$ decomposes in $L^2(\mu)$
into antisymmetric and symmetric parts with 
$\mathcal{L}_A=\vs\cdot\nabla$ and  $\mathcal{L}_S=\ps^{-1}\nabla\cdot(\ps D\nabla)$,
where $\mathcal{L}_S$ 
generates 
reversible dynamics 
with 
gradient drift $b_{\mathrm{rev}}(x)=D\nabla\ln\ps(x)$, yielding
the same invariant density
$\ps(x)$ \cite{qian2013decomposition,DaCosta2023}.
We 
assume the dynamics to satisfy a Poincar\'e inequality \cite{bakry2014analysis,pavliotis2014stochastic}, 
a standard assumption for confining dynamics
\footnote{Sufficient conditions for a Poincar\'e inequality are, e.g., 
the Bakry-\'Emery curvature condition $\nabla^2\Phi\ge\kappa I$ for isotropic noise
$D=I$ and $\ps \propto\e^{-\Phi}$, which yields $\lambda_\mathrm{gap}\ge\kappa$
\cite{bakry2014analysis,pavliotis2014stochastic}
or Lyapunov drift conditions \cite{bakry2008rate,down1995exponential,meyn1993stability}.
}, 
\begin{align}
    \Var_\mu(f) \leq - C_{\mathrm{P}} \langle \mathcal{L}f,f\rangle_\mu,
\end{align}
for some finite $C_{\mathrm{P}}>0$ with 
best constant
$C_{\mathrm{P}}=1/\lambda_{\mathrm{gap}}$, 
with the spectral gap~\footnote{The spectral gap can in principle be estimated from measured time-series via a lower bound \cite{Donsker1975,Donsker1976, Lu2017} or variational upper bounds \cite{vo2025inverse}.}
\begin{align}
    \lambda_{\mathrm{gap}} = \inf_{f:\E^\mu[f]=0}\frac{-\langle \mathcal{L}f,f\rangle_\mu}{\Var_\mu(f)} > 0.
\end{align}
Since
$\langle\mathcal{L}f,f\rangle_\mu=\langle\mathcal{L}_S f,f\rangle_\mu$ even
for irreversible dynamics ($\mathcal{L}_A\neq0$), the spectral gap provides
a variational characterization of the smallest nonzero eigenvalue of
$-\mathcal{L}_S$ and 
sets the slowest timescale of the equilibrium
relaxation. 
At the same time, it guarantees exponential decay of stationary
correlations at rate $\lambda_{\mathrm{gap}}$ also for the irreversible
dynamics, since the antisymmetric part $\mathcal{L}_A$ can only accelerate
relaxation in 
$L^2(\mu)$~\cite{hwang1993accelerating,lelievre2013optimal,duncan2016variance,ReyBellet2016,ReyBellet2015,Duncan2017,Abdulle2019}.

\begin{figure}[tbp]
    \centering
    \includegraphics{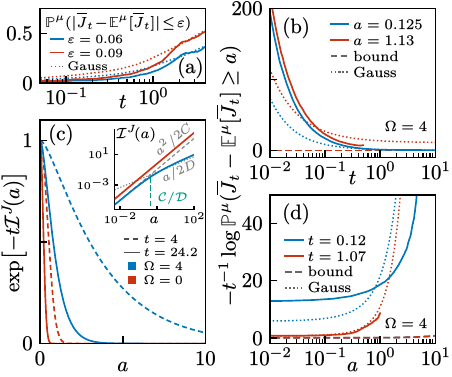}
    \caption{
    (a) Probability that $|\overline{J}_t-\E^\mu[\overline{J}_t]|$
    is
    less than $\varepsilon=0.06$ ($20\%$, blue) or
    $\varepsilon=0.09$ ($30\%$, red) of the stationary mean
    remains small beyond the relaxation time $\lambda_{\mathrm{gap}}^{-1}=1$,
    and is overestimated by the Gaussian approximation.
    (b,d) Empirical rate 
    $-t^{-1}\log\mathbb{P}^\mu(\overline{J}_t-\E^\mu[\overline{J}_t]\ge a)$
    compared with $\mathcal{I}^J(a)$ of the bound
    and a Gaussian approximation, as a function of (b) $t$ at fixed $a$ 
    and (d) $a$ at fixed $t$.
    (c) Concentration bound as a function of $a$ at different times for the 
    driven (blue) and equilibrium (red) dynamics.
    Inset: 
    Interpolation between the Gaussian regime
    $a^2/2\mathcal{C}$ at small and the exponential regime 
    $a/2\mathcal{D}$ at large deviations, crossing over at $a^\ast=\mathcal{C}/\mathcal{D}$
    (green); at equilibrium (red) it remains Gaussian at all $a$.
    Panels show the two-dimensional Ornstein-Uhlenbeck process ($\Omega=4$),
    $V_z^h$ at $z=(1,0)$ with $h=1$ along $\hat n=(0,-1)$ and $\lambda_\mathrm{gap}=1$.
    }
    \label{fig:fig2}
\end{figure}

We consider time-averaged generalized currents, i.e., 
additive functionals of Stratonovich type
\begin{align}
    \overline{J}_t(U) = \frac{1}{t} \int_0^t U(X_s) \circ dX_s,
    \label{eq:schdrom}
\end{align}
where $U$ is a bounded differentiable vector field and $\circ$ denotes the Stratonovich convention with
$U(X_s)\circ d X_s=\sum_i U_i(X_s)\circ d X^i_s$. The observable $U$ has \emph{intrinsic local fluctuations} $Q\equiv U\cdot DU$.
For example, given a 
window function
$V_z^h\geq0$
centered at $z$ with $\int_{\mathbb{R}^d} V^h_z dx=1$,
the choice $U=V^h_z \hat n$ (with $\hat n \in \mathbb{R}^d$ and $\|\hat{n} \|=1$)
yields 
a coarse-grained estimator $\overline{J}_t(V_z^h \hat n)$ of the local steady-state current $\js$
along $\hat n$ at $z$ with scale $h$ \cite{dieball2022mathematical,dieball2022coarse}
(see Fig.~\ref{fig:fig1}c and SM \cite{SM}).
To exemplify the need for uncertainty bounds in Eq.~\eqref{eq:conc_intro}
we show in Fig.~\ref{fig:fig2}a that the probability that
$\overline{J}_t(U)/\E^\mu[\overline{J}_t]-1\in[-\varepsilon,\varepsilon]$
with $\varepsilon\in\{0.2, 0.3\}$ is low even for $t\geq\lambda^{-1}_\mathrm{gap}$
for the driven OUP.

The observable only 
probes the region 
over 
$\mathrm{supp}(U)\equiv\{x\in\mathbb{R}^d:U(x)\neq0\}$, 
so that, for irreversible dynamics, the relevant
dissipation is the \emph{observed} mean steady-state entropy production rate
(in units of $k_\mathrm{B}$)~\cite{seifert2012stochastic,seifert2025stochastic,DaCosta2023}
\begin{align}
  \!\Sigma^U \!\equiv\! \int_{\mathrm{supp}(U)}\!\!\!
  \!\!\!\!\!\!\!\!\!\vs(x)\cdot D^{-1} \vs(x) \ps(x) d x
  = \langle \Sigma_\mathrm{loc}\mathbb{1}_{\mathrm{supp}(U)}\rangle_\mu,
\end{align}
where $\Sigma_\mathrm{loc}(x)\equiv\vs(x)\cdot D^{-1}\vs(x)\geq 0$ 
denotes the local (i.e., pointwise) steady-state entropy production rate. 
Its supremum over the probed region defines the \emph{observed} maximal dissipation rate
$\Sigma^U_\infty\equiv\|\Sigma_\mathrm{loc}\mathbb{1}_{\mathrm{supp}(U)}\|_{L^\infty(\mu)}$, and \emph{global} rates are recovered for a fully supported observable,
$\Sigma\equiv\langle\Sigma_\mathrm{loc}\rangle_\mu\geq\Sigma^U$ and
$\Sigma_\infty\equiv\|\Sigma_\mathrm{loc}\|_{L^\infty(\mu)}\geq\Sigma^U_\infty$.

\emph{Thermodynamic concentration inequalities.---}We
prove upper bounds using the Cram\'er-Chernoff 
approach~\cite{boucheron2013concentration} together with the Lumer-Phillips theorem \cite{engel2000one}. The required control of the moment-generating function is established
by controlling the principal eigenvalue of the tilted 
generator~\cite{dieball2023feynman}
through its symmetrization and using the Poincar\'e inequality 
on $\mathcal{L}_S$ 
(see~\cite{SM,Arxiv_Stratonovich}).

For any generalized current $\overline{J}_t(U)$ 
and any duration $t$, the \emph{thermodynamic concentration inequality} holds
\begin{align}
  \mathbb{P}^\nu\!\bigl(\overline{J}_t-\E^\mu[\overline{J}_t]\geq a\bigr)
  \leq N_\nu\exp\left[-\frac{t\mathcal{C}}{\mathcal{D}^{2}}
  h\!\left(\frac{\mathcal{D}a}{\mathcal{C}}\right)\right],
  \label{Schdabbil}
\end{align}
i.e., $\mathcal{I}^J(a)\equiv(\mathcal{C}/\mathcal{D}^2) h(\mathcal{D}a/\mathcal{C})$
where $h(u)\equiv1+u-\sqrt{1+2u}$ and 
$N_\nu\equiv\left\|d\nu/d\mu\right\|_{L^2(\mu)}$ encodes
the initial conditions ($N_\nu=1$ if $\nu=\mu$).
The 
bound has a characteristic sub-gamma structure \cite{boucheron2013concentration} interpolating between Gaussian and exponential regimes (see End Matter)
and the underlying dynamics only enter via the constants
\begin{align}
    \mathcal{C}\equiv 
     2\|Q\|_{L^\infty(\mu)}\left(1 
    + \frac{\Sigma^U}{\lambda_\mathrm{gap}}\right),
\quad
    \mathcal{D}\equiv
    \frac{2\sqrt{\Sigma^U_\infty\|Q\|_{L^\infty(\mu)}}}{\lambda_\mathrm{gap}}.
    \label{eq:Konschdanden}
\end{align}
Since $\Sigma^U\leq \Sigma$ and $\Sigma^U_\infty\leq\Sigma_\infty$,
Eq.~\eqref{Schdabbil} 
also holds with global dissipation rates,
and further 
holds 
for the left tail $\mathbb{P}^\nu(\E^\mu[\overline{J}_t]-\overline{J}_t\geq a)$ 
since $U\to -U$ leaves $\mathcal{C},\mathcal{D}$ invariant. Moreover, a union bound 
yields the two-sided statement at the cost of an additional 
factor of $2$, $\mathbb{P}^\nu(|\overline{J}_t-\E^\mu[\overline{J}_t]|\geq a)\leq 2 N_\nu \exp[-t\mathcal{I}^J(a)]$.

At detailed balance ($\js=0$), 
$\Sigma^U=\Sigma^U_\infty=0$
and the stationary mean of any
generalized current vanishes 
identically~\footnote{
$\E^\mu[\overline{J}_t]=\int_{\mathbb{R}^d}U(x)\cdot \js(x) dx$
for Stratonovich-type currents \cite{dieball2022mathematical}}, yet
fluctuations persist.
Our bounds then give $\mathcal{C}^{\mathrm{eq}}=2\|Q\|_{L^\infty(\mu)}$
and $\mathcal{D}^\mathrm{eq}=0$, rendering 
fluctuations purely sub-Gaussian, $\exp[-ta^2/(2\mathcal{C}^\mathrm{eq})]$, at \emph{all} deviation scales.
Out of equilibrium ($\js\ne 0$), dissipation enters \emph{quantitatively},  
widening the Gaussian scale by a factor of $(1+\Sigma^U/\lambda_{\mathrm{gap}})$, 
and a \emph{qualitatively new}, exponential scale 
emerges which, beyond a crossover 
of $a^\ast\equiv\mathcal{C}/\mathcal{D}$,
dominates and leads to heavier-than-Gaussian tails for large deviations.
Broken detailed balance thus leaves a fingerprint on the 
shape of the fluctuations
of \emph{any}
current.

\begin{figure}[tbp]
    \centering
    \includegraphics{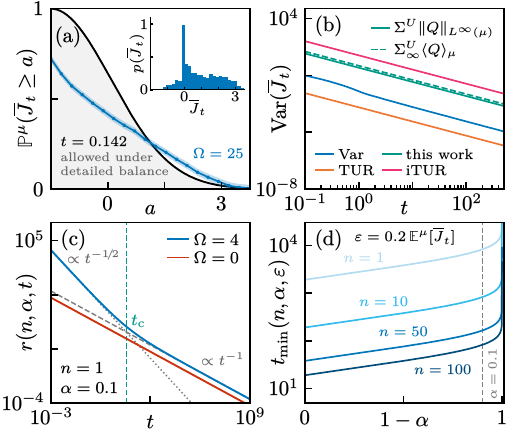}
    \caption{
    (a) 
    Violation of the equilibrium bound (grey)
    by the empirical deviation probability (blue) 
    for driven dynamics ($\Omega=25$) detects irreversibility even at short $t$.
    Inset: Histogram of $\overline{J}_t$.
    (b) Steady-state variance $\Var_\mu(\overline{J}_t)$ as a function of $t$,
    bounded from below by the TUR (orange) and from above by the existing 
    iTUR (magenta) and our refined result (green).
    (c) Confidence radius $r(n,t, \alpha)$ for
    $n=1$ and $\alpha=0.1$  as function of $t$.
    Out of equilibrium (blue), the tail contribution $\propto t^{-1}$
    dominates the error bars up to the crossover time $t_\mathrm{c}$ (green),
    beyond which Gaussian scaling $\propto t^{-1/2}$ takes over.
    At equilibrium (red), error bars are Gaussian at all times.
    (d) Minimal observation time $\tmin(n,\alpha,\varepsilon)$ guaranteeing 
    an accuracy $\varepsilon=0.2\E^\mu[\overline{J}_t]$ 
    (i.e., relative error $\pm 20\%$) as a function of the confidence level $1-\alpha$,
    for different sample sizes $n$.
    All panels show the two-dimensional Ornstein-Uhlenbeck process with 
    (a)
    $V_z^h$ at $z=(0.5,0)$ with $h=1.5$ along $\hat n=(0,-1)$
    and
    (b-d)
    $V_z^h$ at $z=(1,0)$ with $h=1$ along $\hat n=(0,-1)$.
    }
   \label{fig:fig3}
\end{figure}

We leverage this 
to construct a 
model-free, 
nonasymptotic test for 
broken detailed balance (see App.~F).
Since detailed balance
enforces 
sub-Gaussian fluctuations for \emph{any}
generalized current, statistically significant deviations 
from 
this bound (see Fig.~\ref{fig:fig3}a) 
rigorously reveal irreversibility
from operationally accessible quantities, 
without a reference
measurement 
or knowledge of the driving, the dissipation rate, or relaxation timescale. See also \cite{Thapa2021} for 
results on detecting deviations from Brownian motion.   

\emph{Dissipation concentration inequalities.---}For suitable 
$U$, 
$\overline{J}_t$
realizes generalized currents central to stochastic thermodynamics,
such that our concentration inequalities 
bound the
full distribution of 
work, heat, or entropy
production.
For example, for a smooth window function $0\leq K_z^h\leq 1$ localized around $z$ at scale $h$,
$U_\Sigma=D^{-1}\vs K^h_z$---with $\mathrm{supp}(U_\Sigma)=\mathrm{supp}(K^h_z)$
defining the observed region---yields the current
$\Sigmaline^K\equiv\overline{J}_t\big(D^{-1}\vs K^h_z\big)$, i.e.,
a 
trajectory estimator of the coarse-grained regional entropy production rate
$\E^\mu[\Sigmaline^K]=\langle\Sigma_\mathrm{loc}K^h_z\rangle_\mu\equiv\Sigma^K\leq\Sigma^U$.

Here $\Sigma^K$ is the coarse-grained 
and $\Sigma^U$ is the full
mean entropy production rate 
of the probed region. 
The two coincide in the limit $K_z^h\to\mathbb{1}_{\mathrm{supp}(U_\Sigma)}$.
In analogy 
to 
coarse-grained density and current estimators \cite{dieball2022mathematical, dieball2022coarse}, 
$\Sigmaline^K$
thus resolves dissipation at scale $h$.
Then, for any $t>0$ the \emph{dissipation concentration inequality} holds
\begin{align}
  \mathbb{P}^\nu\bigl(\Sigmaline^K-\Sigma^K\ge a\bigr)
  \le N_\nu\exp\left[-\frac{t\mathcal{C}_\Sigma}{\mathcal{D}_\Sigma^{2}}
  h\left(\frac{\mathcal{D}_\Sigma a}{\mathcal{C}_\Sigma}\right)\right],
  \label{eq:diss_conc_ineq}
\end{align}
with constants 
that encode $\Sigma^U$
and $\lambda_\mathrm{gap}$ 
\begin{align}
  \mathcal{C}_\Sigma
  = 2\Sigma^U_\infty\left(1+\frac{\Sigma^U}{\lambda_\mathrm{gap}}\right),
  \qquad
  \mathcal{D}_\Sigma
  = \frac{\Sigma^U_\infty}{\lambda_\mathrm{gap}}.
  \label{eq:diss_constants}
\end{align}
Thus, the locally
observed dissipation bounds the fluctuations of its own estimator, and 
the dynamics enter only through 
$\Sigma^U$, $\Sigma^U_\infty$, and
$\lambda_\mathrm{gap}$. 
For a fully supported window function $K_z^h$ and bounded maximal rate $\Sigma_\infty<\infty$,
we recover a global statement via $\Sigma^K,\Sigma^U\to\Sigma$ and $\Sigma^U_\infty\to\Sigma_\infty$.
Left-tail and two-sided statements follow
as before.

\emph{Inverse thermodynamic uncertainty relation.---}%
Beyond tails, 
the concentration-of-measure viewpoint 
bounds the steady-state variance of any generalized current
from above. 
Expanding the bound on the moment-generating function 
(see Proposition 5.1 in \cite{Arxiv_Stratonovich}) %
to second order yields, for all $t>0$, the \emph{refined} inverse thermodynamic uncertainty relation (iTUR) 
(see Fig.~\ref{fig:fig3}b and App.~E),
\begin{align}
  \mathrm{Var}_\mu(\overline{J}_t)
  &\leq
  \frac{2}{t}  
  \left(  
  \langle Q\rangle_\mu
  + \frac{\langle\Sigma^U Q\rangle^{\infty}_{\wedge\vee}
    - \E^\mu[\overline{J}_t]^2}{\lambda_\mathrm{gap}}
  \right),
  \label{eq:inverse_turL}
\end{align}
where $\langle\Sigma^U Q\rangle^{\infty}_{\wedge\vee}\equiv \min\{\Sigma^U_\infty\langle Q\rangle_\mu, \Sigma^U \|Q\|_{L^\infty(\mu)} \}$.
In contrast to existing results, \emph{only}
the dissipation observed by $U$ enters,
and using the weaker bound $\langle\Sigma^U Q\rangle^{\infty}_{\wedge\vee}\leq\Sigma \|Q \|_{L^\infty(\mu)}$
recovers the recent iTUR for overdamped Langevin dynamics 
\cite{vo2025inverse} up to a factor
(see App.~E).
The refinement 
is most prominent whenever the current probes
weakly dissipative regions (i.e., $\Sigma^U\ll\Sigma$).
Local parameters emerge
by the concentration 
pointwise controlling the full distribution,
which requires bounding 
the tilted generator via explicit control of the local 
quantities $W=\vs \cdot U$ and $Q=U\cdot D U$ \cite{Arxiv_Stratonovich}, 
both of which vanish outside $\mathrm{supp}(U)$.
Together with the TUR, Eq.~\eqref{eq:inverse_turL}
thus gives a sandwich bound for current fluctuations.

\emph{Thermodynamic uncertainty quantification.---}
Since the Cram\'er-Chernoff method 
relies on 
bounding 
the moment-generating function of $\overline{J}_t$,
the single-trajectory bounds in Eq.~\eqref{Schdabbil} 
extend to the sample mean $\widehat{J}_{n,t}$
over independent realizations
upon the substitutions $t\to nt$ and $N_\nu\to N_\nu^{n}$ 
(see End Matter and \cite{boucheron2013concentration,bebon2023controlling,Arxiv_Stratonovich}).
This yields
nonasymptotic (i.e., finite-time and small-sample) performance guarantees.
For 
any generalized current, 
$t>0$, and $n\geq 1$, with probability of at least $1-\alpha$ under $\mathbb{P}^\nu$ 
it holds 
that 
    $\E^\mu[\overline{J}_t]\in [\widehat{J}_{n,t}- r, \widehat{J}_{n,t} +r ]$ (Fig.~\ref{fig:fig1}b), with
\begin{align}
    r(n,t,\alpha)\equiv 
    \sqrt{\frac{2\mathcal{C}}{nt}\log(2N_\nu^{n}/\alpha)}
    + \frac{\mathcal{D}}{nt}\log(2N_\nu^{n}/\alpha)
    ,
    \label{eq:conf_radius}
\end{align}
denoting the explicit confidence radius.
Notably, $r$ does \emph{not} vanish as $n\to\infty$ at fixed $t$ unless $\nu=\mu$, i.e., 
more trajectories cannot compensate for nonstationary initial conditions (see App.~D).

Qualitative 
and quantitative
differences between reversible and irreversible 
systems 
manifest 
directly in the structure of the confidence radius 
\eqref{eq:conf_radius}.
Under detailed balance the tail scale vanishes, $\mathcal{D}^\mathrm{eq}=0$,
and the radius reduces to the first term with purely Gaussian scaling $r\propto(nt)^{-1/2}$,
i.e., 
equilibrium 
currents have \emph{(sub-) Gaussian errors at all times}.
Out of equilibrium, 
dissipation acts \emph{quantitatively}
via $\mathcal{C}$, which carries the factor $(1+\Sigma^U/\lambda_\mathrm{gap})$ 
and increases
the Gaussian contribution 
at all times.
Moreover, 
it contributes \emph{qualitatively} through
the second term 
which 
covers
the exponential (heavier-than-Gaussian) tails of $\overline{J}_t$ 
that 
nonasymptotic confidence intervals
must account for at finite times.
This non-Gaussian correction decays as $(nt)^{-1}$
and is thus subleading at long times,
but dominates 
prior to the crossover time (see Fig.~\ref{fig:fig3}c)
\begin{align}
    t_c(n,\alpha) \equiv \frac{\Sigma^U_\infty}{\lambda_\mathrm{gap}(\lambda_\mathrm{gap}+\Sigma^U)}
    \frac{\log(2N_\nu^{n}/\alpha)}{n}.
\end{align}
Since $t_c\lambda_\mathrm{gap}\propto\Sigma^U_\infty/(\lambda_\mathrm{gap}+\Sigma^U)$,
at fixed $n$,
the 
non-Gaussian regime $[0,t_c]$
extends over many relaxation times whenever 
$\Sigma^U_\infty$ exceeds both
$\lambda_\mathrm{gap}$ and $\Sigma^U$, i.e., 
for strongly heterogeneous driving in the observed region.

Inverting $r\leq\varepsilon$ 
for a prescribed tolerance $\varepsilon$
at fixed sample size $n$ 
yields
$\mathbb{P}^\nu(-\varepsilon \leq \widehat{J}_{n,t} - \E^\mu[\overline{J}_t]\leq \varepsilon)\geq1-\alpha$, for all $t\geq \tmin$ with $\Theta(\varepsilon)\equiv(\sqrt{\mathcal{C}}+\sqrt{\mathcal{C}+2\mathcal{D}\varepsilon})^{2}/2\varepsilon^{2}$ and
\begin{align}
  \tmin(n,\alpha,\varepsilon)\equiv
  \frac{\Theta(\varepsilon)}{n}\log(2N_\nu^{n}/\alpha).
  \label{time}
\end{align}
Equivalently, at fixed trajectory length
$t>\Theta(\varepsilon)\log N_\nu$ the statement holds for every sample size $n\geq\nmin$ with
\begin{align}
  \nmin(t,\alpha,\varepsilon)\equiv
  \left\lceil\frac{\Theta(\varepsilon)\log(2/\alpha)}
  {t-\Theta(\varepsilon)\log N_\nu}\right\rceil.
  \label{samples}
\end{align}
Since 
$\mathcal{C}=\mathcal{C}^{\mathrm{eq}}(1+\Sigma^U/\lambda_\mathrm{gap})$ with
the equilibrium coefficient
$\mathcal{C}^{\mathrm{eq}}=2\|Q\|_{L^\infty(\mu)}$, the sandwich
bound $2\mathcal{C}/\varepsilon^{2}\leq\Theta(\varepsilon)\leq
2\mathcal{C}/\varepsilon^{2}+2\mathcal{D}/\varepsilon$ yields the floors
\begin{align}
  \tmin(n,\alpha,\varepsilon)
  &\geq
  \left(1+\frac{\Sigma^U}{\lambda_\mathrm{gap}}\right)
  \frac{\Theta^\mathrm{eq}(\varepsilon)}{n}\log(2N_\nu^{n}/\alpha),
  \nonumber\\
  \nmin(t,\alpha,\varepsilon)
  &\geq
  \left\lceil\frac{(1+\Sigma^U/\lambda_\mathrm{gap})
  \Theta^\mathrm{eq}(\varepsilon)\log(2/\alpha)}
  {t-(1+\Sigma^U/\lambda_\mathrm{gap})
  \Theta^\mathrm{eq}(\varepsilon)\log N_\nu}\right\rceil,
  \label{eq:tmin_nmin}
\end{align}
with $\Theta^\mathrm{eq}(\varepsilon)\equiv2\mathcal{C}^{\mathrm{eq}}/\varepsilon^{2}$.
The minimal observation time and sample size for irreversible dynamics
exceed 
their equilibrium counterparts by a 
dissipation factor, i.e.,
$\tmin\geq(1+\Sigma^U/\lambda_\mathrm{gap})\tmin^{\mathrm{eq}}$. 
For nonstationary initial
conditions ($N_\nu>1$), dissipation also enters the 
threshold 
$t>(1+\Sigma^U/\lambda_\mathrm{gap})\Theta^\mathrm{eq}(\varepsilon)\log N_\nu$,
below which \emph{no} number of independent 
measurements 
can compensate for trajectories that are too short. Moreover, 
higher confidence is 
cheap due to the 
dependence $\propto \log(1/\alpha)$
(see Fig.~\ref{fig:fig3}d). 

The crucial implication of all results is \emph{not} that central-limit 
approximations 
can become inaccurate; unlike our bounds \emph{they can be violated} 
(see Fig.~\ref{fig:fig2}b,d),
and by
significantly underestimating tail probabilities,
confidence guarantees based on these arguments 
have less coverage than required and fail to be valid.

\emph{Conclusion.---}Leveraging the Cram\'er-Chernoff method 
we derived nonasymptotic bounds on the probability that an observable generalized current $\overline{J}_t$, or its sample mean over any arbitrary sample size $n$, of a geometrically ergodic diffusion deviates from the stationary mean $\E^\mu[\overline{J}_t]$ by more than any amount $a$. 
The concentration 
is governed by the relaxation time of the diffusive dynamics, the locally observed dissipation rate, and the intrinsic local fluctuations of the observable, and
reveals qualitative and quantitative differences in 
fluctuations in and out of thermodynamic equilibrium. 
We further obtained \emph{refined} inverse thermodynamic uncertainty relations, 
bounding 
generalized current variances
from above,
and
constructed 
nonasymptotic confidence intervals enabling uncertainty control 
in thermodynamic inference 
from short trajectories and small samples. 
Importantly, Eqs.~\eqref{time} and~\eqref{samples} provide the first 
\emph{quantitative answer} to when a trajectory is sufficiently long and a sample is sufficiently large.
We applied our results to currents observed on a two-dimensional
Ornstein-Uhlenbeck process in and out of equilibrium 
and 
rigorously detected broken detailed balance. 
Further practical applications 
and results for sub-geometrically ergodic and underdamped (phase-space) and multiplicative noise \cite{ZimmerArxiv} diffusions, 
and Markov-jump processes,
will be addressed in future works.

\emph{Acknowledgments.---} Financial support from the Studienstiftung des Deutschen Volkes (to R.~B.) and from the European Research
Council (ERC) under the European Union’s Horizon Europe
research and innovation program (Grant Agreement
No.~101086182 to A.~G.) 
is gratefully acknowledged. 

\bibliographystyle{apsrev4-2.bst}
\bibliography{bibliography}

\newpage
\onecolumngrid
\section*{End Matter}
\twocolumngrid
\renewcommand{\theequation}{A\arabic{equation}}
\setcounter{equation}{0} 

\emph{Appendix~A: Proof sketch.---}%
The thermodynamic concentration inequality~\eqref{Schdabbil} 
follows from a
general concentration inequality for additive functionals of Stratonovich
type~\eqref{eq:schdrom}, 
which we recently proved in~\cite{Arxiv_Stratonovich} 
(see also~\cite{SM} for the proof strategy),
\begin{align}
    \mathbb{P}^\nu(\overline{J}_t - \E^\mu[\overline{J}_t] \geq a)
    \leq N_\nu \exp\left[-\frac{t\widetilde{\mathcal{C}}}{\widetilde{\mathcal{D}}^{2}}
  h\left(\frac{\widetilde{\mathcal{D}}a}{\widetilde{\mathcal{C}}}\right)\right],
  \label{eq:aap_bound}
\end{align}
where, with $W\equiv \vs \cdot U$ and $Q\equiv U\cdot D U$,
\begin{align}
    \widetilde{\mathcal{C}}
    \equiv 2 \| Q \|_{L^\infty(\mu)} + \frac{2\Var_\mu(W)}{\lambda_\mathrm{gap}},
    \quad
    \widetilde{\mathcal{D}}
    \equiv \frac{\| W-\langle W\rangle_\mu \|_{L^\infty(\mu)}}{\lambda_\mathrm{gap}}.
    \label{eq:aap_constants}
\end{align}
These general constants do not provide any direct physical
insights and the thermodynamic
content of Eq.~\eqref{Schdabbil}
enters when they are bounded by the observed dissipation rates.
Since Eq.~\eqref{eq:aap_bound}
is increasing in both parameters, it suffices to bound them from above.
First, $\widetilde{\mathcal{C}}\leq\mathcal{C}$ with $\mathcal{C}$ from
Eq.~\eqref{eq:Konschdanden} follows from
$\Var_\mu(W)\leq \langle\Sigma^U Q\rangle^{\infty}_{\wedge\vee}
- \E^\mu[\overline{J}_t]^2\leq \langle\Sigma^U Q\rangle^{\infty}_{\wedge\vee}
\leq \Sigma^U \| Q \|_{L^\infty(\mu)}$.
Second, $\widetilde{\mathcal{D}}\leq\mathcal{D}$ follows from
$\| W-\langle W\rangle_\mu \|_{L^\infty(\mu)}
\leq \sqrt{\Sigma^U_\infty\|Q\|_{L^\infty(\mu)}}
+ \sqrt{\Sigma^U\langle Q\rangle_\mu}
\leq 2 \sqrt{\Sigma^U_\infty\|Q\|_{L^\infty(\mu)}}$ (see~\cite{SM}).

\emph{Appendix B: Additional thermodynamic bounds.---}%
The parameters~\eqref{eq:Konschdanden} of the main text
use the respective last inequality in Appendix~A, which is the simplest (and easiest to interpret)
but more conservative.
Using a sharper inequality gives
sharper bounds so that Eq.~\eqref{Schdabbil}
also holds with 
\begin{align}
    \widetilde{\mathcal{C}}
    \leq \mathcal{C}' &\equiv 2\left(\| Q\|_{L^\infty(\mu)}
    + \frac{\langle \Sigma^U Q\rangle^\infty_{\wedge\vee}
    -\E^\mu[\overline{J}_t]^2}{\lambda_\mathrm{gap}}\right)\leq \mathcal{C},
    \nonumber
    \\
    \widetilde{\mathcal{D}}\leq \mathcal{D}' &\equiv
    \frac{\sqrt{\Sigma^U_\infty\|Q\|_{L^\infty(\mu)}}
    + \sqrt{\Sigma^U\langle Q\rangle_\mu}}{\lambda_\mathrm{gap}}
    \leq \mathcal{D}.
    \label{eq:sharp_constants}
\end{align}

Moreover, we use a second independent approach 
in \cite{Arxiv_Stratonovich}, which 
gives Eq.~\eqref{eq:aap_bound} with 
different general constants
$\widetilde{\mathcal{C}}_2$, $\widetilde{\mathcal{D}}_2$
that include $W$ and $Q$.
In the same fashion these can again be bounded in terms of
thermodynamic quantities, then giving
different constants 
$\mathcal{C}_2$, $\mathcal{D}_2$ 
for Eq.~\eqref{Schdabbil}
(see~\cite{SM}).

\emph{Appendix C: Bernstein bounds.---}%
The thermodynamic concentration bound~\eqref{Schdabbil}
has a characteristic sub-gamma structure \cite{boucheron2013concentration}
that, using the elementary inequality $h(u)=1+u -\sqrt{1+2u} \geq u^2/[2(1+u)]$ for $u\geq0$, can be relaxed to the simpler \emph{Bernstein inequality}
\begin{align}
    \exp\left[-\frac{t\mathcal{C}}{\mathcal{D}^{2}}
  h\left(\frac{\mathcal{D} a}{\mathcal{C}}\right)\right]
  \leq
  \exp\left[-\frac{ta^2}{2(\mathcal{C}+\mathcal{D} a)}\right].
\end{align}
This highlights that for small 
deviations $a\ll \mathcal{C}/\mathcal{D}$
the bound is \emph{sub-Gaussian} with a variance proxy of
$\mathcal{C}/t$, while for large deviations $a\gg \mathcal{C}/\mathcal{D}$ 
it decays only exponentially at a rate $t/2\mathcal{D}$,
the crossover occurring at $a=\mathcal{C}/\mathcal{D}$ (see Fig.~\ref{fig:fig2}c).
The same 
holds for the 
dissipation concentration inequality~\eqref{eq:diss_conc_ineq}
with $\mathcal{C}\to\mathcal{C}_\Sigma$ and
$\mathcal{D}\to\mathcal{D}_\Sigma$.

\emph{Appendix~D: Nonstationary initial condition and sample size.---}%
At fixed trajectory length $t$,
the confidence radius~\eqref{eq:conf_radius}
saturates as $n\to\infty$,
\begin{align}
    r_\infty\equiv\lim_{n\to\infty} r(n,t,\alpha)
    = \sqrt{\frac{2\mathcal{C}\log (N_\nu)}{t}} + \frac{\mathcal{D}\log(N_\nu)}{t}.
    \label{eq:r_floor}
\end{align}
Increasing the sample
size thus cannot compensate for nonstationary initial conditions ($N_\nu >1$).
Note that $N_\nu=1$
if and only if $\nu=\mu$, in which case
$r_\infty=0$
for any finite $t$.
Requiring
$r_\infty\leq\varepsilon$
recovers the threshold $t\geq\Theta(\varepsilon)\log N_\nu$
stated below Eq.~\eqref{eq:tmin_nmin}.
Equivalently, 
the crossover time $t_c$ after 
which error bars become Gaussian
tends to $t_c\to\Sigma^U_\infty\log N_\nu/[\lambda_\mathrm{gap}(\lambda_\mathrm{gap}+\Sigma^U)]$
as $n\to\infty$, i.e., even for asymptotic sample sizes
the non-Gaussian regime $[0,t_c]$ vanishes only when
$\nu=\mu$.

\emph{Appendix E: Additional inverse thermodynamic uncertainty relations.---}%
Using $\langle Q\rangle_\mu\leq\|Q\|_{L^\infty(\mu)}$ in
Eq.~\eqref{eq:inverse_turL} gives the weaker but simpler bounds
\begin{align}
  \Var_\mu(\overline{J}_t)
  &\leq
  \frac{2}{t}
  \left(
  \|Q\|_{L^\infty(\mu)}
  + \frac{\langle\Sigma^U Q\rangle^{\infty}_{\wedge\vee}
    - \E^\mu[\overline{J}_t]^2}{\lambda_\mathrm{gap}}
  \right)
 \nonumber  
 \\  
 &\leq
  \frac{2}{t}
  \|Q\|_{L^\infty(\mu)}  
  \left(1+\frac{\Sigma^U}{\lambda_\mathrm{gap}}
     \right)
  \label{eq:itur_infty2}
\end{align}
where the last inequality follows from $\langle \Sigma^UQ\rangle_{\wedge\vee}^\infty\leq \Sigma^U\|Q\|_{L^\infty(\mu)}$
and $\E^\mu[\overline{J}_t]^2\geq 0$.
The right-hand sides correspond 
to the sub-Gaussian variance proxies $\mathcal{C}'/t$ and $\mathcal{C}/t$, respectively, 
which are, by definition, upper bounds on the variance \cite{boucheron2013concentration}.

Our iTUR can be considered a refinement since
first, the dissipation only enters via
observed quantities $\Sigma^U, \Sigma_\infty^U$, 
and the minimum
in $\langle\Sigma^U Q\rangle^{\infty}_{\wedge\vee}$
selects the better case depending on
the inhomogeneity of either $Q$ or $\Sigma^U$.
When relaxing both---i.e., $\Sigma^U\to\Sigma$
and keeping only the second term in 
the minimum---gives
the term $\Sigma\|Q\|_{L^\infty(\mu)}$ of \cite{vo2025inverse}
up to a factor of
$g(\bar\lambda_\mathrm{gap}t)\leq1$, with
$\bar\lambda_\mathrm{gap}\equiv\sqrt{\lambda_\mathrm{gap}
(\lambda_\mathrm{gap}+\Sigma_\infty)}$, increasing monotonically from $0$ to $1$.
Since $g\to1$ as $t\to\infty$,
and also at finite $t$ for unbounded processes like the OUP where
$\Sigma_\infty\to\infty$,
the factor gives no sharper bound in these regimes even when $\Sigma^U\approx \Sigma$.
In principle, for 
$\Sigma_\infty<\infty$ and small $t$, however, $g$
can become small enough that
the bound becomes tighter than Eq.~\eqref{eq:inverse_turL}
despite the local refinement.
We therefore conjecture that
combining our local approach with the proof of \cite{vo2025inverse}
gives Eq.~\eqref{eq:inverse_turL} with
$g(\bar\lambda_\mathrm{gap}t)$ included
which sharpens the bound for finite $t$
in cases when $\Sigma_\infty < \infty$.

\emph{Appendix F: Detecting broken detailed balance.---}%
The equilibrium form of our concentration bound~\eqref{Schdabbil}
yields a simple, nonasymptotic test of whether 
the fluctuations of \emph{any}
inferred current
are consistent with detailed balance. 
Under detailed balance, 
the mean of any generalized
current vanishes, $\E^\mu[\overline{J}_t]=0$, 
and the exponential tail contribution vanishes,
$\mathcal{D}^{\mathrm{eq}}=0$, 
so that Eq.~\eqref{Schdabbil} reduces
to the sub-Gaussian bound
\begin{align}
    \!\!\mathbb{P}^\nu(\overline{J}_t\geq a)\leq N_\nu
    \e^{-t\mathcal{I}^J_{\mathrm{eq}}(a)},
    \quad
    \mathcal{I}^J_{\mathrm{eq}}(a)\equiv\frac{a^2}{4\|Q\|_{L^\infty(\mu)}}.
    \label{eq:eq_bound}
\end{align}
Given $n$ independent trajectories of length $t$,
we test for violations of detailed balance as follows.

\begin{enumerate}
    \item Choose an observable $U$ and compute $\overline{J}_t$
    for each trajectory.
    \item Fix $a$ and count the number of trajectories $n_a$
    that satisfy $\overline{J}_t\geq a$.
    \item Estimate $P^\nu_a\equiv\mathbb{P}^\nu(\overline{J}_t\geq a)$
    by the empirical probability $\hat{P}^\nu_a\equiv n_a/ n$.
    \item If $\hat{P}^\nu_a-\sqrt{\log(1/\alpha)/(2n)}
    \geq N_\nu \exp\bigl[-t\mathcal{I}^J_{\mathrm{eq}}(a)\bigr]$
    for any $a$ for a chosen $U$, we can rule out detailed balance 
    with a confidence of $1-\alpha$.
\end{enumerate}
Notably, the uncertainty of $\hat{P}^\nu_a$
is accounted for by the nonasymptotic lower error bound $\sqrt{\log(1/\alpha)/(2n)}$
from 
Hoeffding's inequality \cite{boucheron2013concentration}
(see \cite{SM}).
Moreover, the equilibrium bound~\eqref{eq:eq_bound}
requires only $\|Q\|_{L^\infty(\mu)}$
with $Q=U\cdot D U$.
Hence it is operationally accessible
since $U$ is an arbitrary admissible vector field
and $D$ is obtainable 
from the short-time behavior of the mean-squared displacement.
By construction, the test is
sufficient but not necessary for irreversibility, i.e.,
violating Eq.~\eqref{eq:eq_bound}
certifies broken detailed balance, 
whereas remaining within
the allowed region
does \emph{not} confirm detailed balance,
since the bound is not tight and the chosen $U$ 
may be blind to the current.


\clearpage
\newpage
\onecolumngrid

\clearpage
\onecolumngrid

\setcounter{secnumdepth}{3}   
\setcounter{tocdepth}{2}

\renewcommand{\thesection}{S\arabic{section}}
\renewcommand{\thefigure}{S\arabic{figure}}
\renewcommand{\thetable}{S\arabic{table}}
\renewcommand{\theequation}{S\arabic{equation}}

\renewcommand{\thesubsection}{\Alph{subsection}}
\renewcommand{\thesubsubsection}{\arabic{subsubsection}}

\setcounter{section}{0}
\setcounter{figure}{0}
\setcounter{table}{0}
\setcounter{equation}{0}
\setcounter{page}{1}

\begin{center}
{\large \textit{Supplemental Material for}}
\\[0.2cm]
\textbf{Thermodynamic Concentration Inequalities:\\ 
Controlling Uncertainty in Finite-Time and Small-Sample Thermodynamic Inference}
\\[0.2cm]
Rick Bebon\,\orcidlink{0000-0003-2187-0008} and Alja\v{z} Godec\,\orcidlink{0000-0003-1888-6666}\\
\small {\textit{Mathematical Physics and Stochastic Dynamics, Faculty of Mathematics and Physics, University of Freiburg}}
\\[0.6cm]
\end{center}

In this Supplemental Material (SM) we present
additional background and
details of the proofs and the numerical methods used
in the Letter. We provide details on the model system,
outline the proof strategy for the general concentration inequality for Stratonovich-type functionals, and give further context on the thermodynamic bounds, uncertainty quantification, and the test for broken detailed balance presented in the Letter.

\tableofcontents
\newpage

\section{Model system and numerics}
In the Letter we exemplify our results by means of a two-dimensional
Ornstein-Uhlenbeck process (OUP) 
with irreversible driving that leaves the
invariant density and the spectral gap unchanged. 
In this section we give further details on the relevant parameters
and give details
on the numerics.

\subsection{Steady-state density and current for the 
driven two-dimensional Ornstein-Uhlenbeck process}
We consider the two-dimensional OUP described by the It\^{o} (Langevin) equation, $X_t\in\mathbb{R}^2$, 
\begin{align}
    d X_t = b(X_t)d t + \sqrt{2D_0}d W_t,
    \qquad b(x) = -Bx,
    \label{eq:OUP}
\end{align}
with isotropic diffusion matrix $D=D_0\mathbb{1}$, $D_0>0$,
and drift matrix
\begin{align}
    B = 
\begin{pmatrix}
r & -\Omega \\
\Omega & r 
\end{pmatrix},
\end{align}
where $r>0$ gives the strength of the confining potential
$V(x)=\tfrac{r\| x\|^2}{2D_0}$, and $\Omega\in\mathbb{R}$ sets
the strength of the nonequilibrium driving.
The corresponding 
steady-state density $\ps$ and steady-state current $\js$ read
\begin{align}
\ps(x)  &= \frac{r}{2\pi D_0}\exp\left(-\frac{rx^\top x}{2D_0}\right),
\label{eq:ps_OUP}
\\
\js(x)  &= [b(x)-D_0\nabla]\ps(x)
= \Omega \ps 
\begin{pmatrix}
x_2\\
-x_1
\end{pmatrix}.
\label{eq:js_OUP}
\end{align}
The corresponding local mean velocity is the
cockwise rotation, i.e., 
$\vs(x)=\Omega(x_2,-x_1)^\top$, 
and $\nabla\cdot\js=0$.
At equilibrium (i.e., detailed balance) $\Omega=0$ and $\js=0$ as expected.
As mentioned in the Letter, the drift can be split into a reversible and irreversible
part, $b(x) = b_\mathrm{rev}(x) + b_{\mathrm{irr}}(x)$,
where $b_\mathrm{rev}(x)=D\nabla\ln\ps(x)=-D\nabla V(x)$
and $b_{\mathrm{irr}}(x)=\vs(x)$.
Notably Eq.~\eqref{eq:ps_OUP}
is independent of the driving $\Omega$, i.e., 
both dynamics share the same $\ps$.

\subsection{Spectral gap of the driven OUP}
The Markov generator for the two-dimensional OUP from Eq.~\eqref{eq:OUP} reads
\begin{align}
    \mathcal{L}
    =
    - \begin{pmatrix}
        r & -\Omega
        \\
        \Omega & r
    \end{pmatrix}
    x\cdot \nabla 
    + D_0 \nabla^2.
\label{eq:generator_oup}
\end{align}
The corresponding eigenvalues
of Eq.~\eqref{eq:generator_oup} are related 
to the distinct eigenvalues of the drift matrix $B$ \cite{SM_Metafune2002},
such that the spectrum set of the operator $\mathcal{L}$
in the Hilbert space $L^2(\mu)$
is given by \cite{SM_Chen2014}
\begin{align}
    \sigma(\mathcal{L}) = \left\{-(m+n)r + i(m-n)\Omega,\quad m,n=0,1,2,\ldots \right\}
\end{align}
with the associated eigenfunctions
expressed by 
the Hermite-Laguerre-It\^{o} polynomials \cite{SM_Chen2014}.
The smallest non-zero eigenvalue therefore
gives
\begin{align}
    \lambda_\mathrm{gap} = r,\qquad C_\mathrm{P}=r^{-1},
\end{align}
which is independent of $\Omega$, i.e., 
the irreversible driving leaves the spectral gap unchanged.

\subsection{Steady-state entropy production rate of the driven OUP}
The local,
the mean,
and the maximal
steady-state entropy production rates for the two-dimensional OUP [Eq.~\eqref{eq:OUP}]
follow with Eqs.~\eqref{eq:ps_OUP} and~\eqref{eq:js_OUP}, respectively, as
\begin{align}
    \Sigmaloc(x)=\vs(x)\cdot D^{-1}\vs(x) = \frac{\Omega^2 \|x\|^2}{D_0},
    \qquad
    \Sigma = \langle\Sigmaloc\rangle_\mu = \frac{2\Omega^2}{r},
    \qquad
    \Sigma_\infty = \|\Sigmaloc\|_{L^\infty(\mu)} = \infty.
    \label{eq:epr}
\end{align}
Due to the confining, but unbounded, quadratic potential $V(x)$ 
the OUP is an example of an
unbounded process with hence 
\emph{unbounded} maximal dissipation rate, but finite mean dissipation rate.
Bounds that involve the global parameter $\Sigma_\infty$
are therefore trivial, whereas the corresponding
\emph{observed} dissipation rates
$\Sigma^U$ and $\Sigma_\infty^U$ remain finite
for any observable $U$ that has compact support.

\subsection{Generalized currents for the OUP}
As an example for a generalized current 
we take the coarse-grained
empirical current \cite{dieball2022coarse, dieball2022mathematical}, i.e., a
trajectory estimator of the local stationary probability current $\js$
along a direction $\hat{n}\in\mathbb{R}^2$ with $\|\hat{n} \|=1$
obtained via the
choice $U=V^h_z \hat n$.
As a normalized window function of scale $h$ centered around $z$
with compact support we take the
two-dimensional Epanechnikov kernel \cite{SM_Hastie2009}
\begin{align}
    V_z^h(x_1,x_2)=
    \begin{cases}
        \dfrac{2}{\pi h^2}\left[1-\dfrac{(x_1-z_1)^2 + (x_2-z_2)^2}{h^2} \right]
        & \qquad (x_1-z_1)^2 + (x_2-z_2)^2 \leq h^2,
        \\[8pt]
        0 &\qquad\text{otherwise},
    \end{cases}
    \label{eq:EpaKernel}
\end{align}
which satisfies $V_z^h\geq 0$ and
$\int_{\mathbb{R}^2}V_z^h =1$.
With that choice we next compute the parameters of Eq.~(10)
that enter our thermodynamic concentration bound~(9).

First, with $Q = U\cdot D U$ and $W=\vs\cdot U$
we obtain, respectively,
\begin{align}
 Q(x) = D_0 \bigl[V_z^h(x)\bigr]^2,
\qquad
W(x) = \Omega V_z^h(x)(n_1 x_2 - n_2 x_1),
\end{align}
where $\hat{n}=(n_1,n_2)^\top$.
For the sup-norm parameters we obtain the analytical expressions
\begin{align}
    \| Q \|_{L^\infty(\mu)}
    = \frac{4D_0}{\pi^2h^4},
    \qquad
    \Sigma^U_\infty = \frac{\Omega^2(\|z\| + h)^2}{D_0} <\infty.
    \label{eq:supnorm_konschdanden}
\end{align}
where the last relation follows since
$\Sigmaloc$ is increasing around $z$ with radius set by $h$, i.e., 
the maximum is reached at the outer point.

For the thermodynamic concentration bound 
and confidence radius of the main text 
it remains to compute $\Sigma^U$, 
which is obtained, for some given system parameters, by solving 
the following integral numerically
\begin{align}
    \Sigma^U
    = \langle \Sigmaloc \mathbb{1}_{\mathrm{supp}(U)}\rangle_\mu
    = \frac{\Omega^{2}r}{2\pi D_0^{2}}
    \int_{z_1-h}^{z_1+h} d x_1
    \int_{z_2-\sqrt{h^{2}-(x_1-z_1)^{2}}}^{z_2+\sqrt{h^{2}-(x_1-z_1)^{2}}} d x_2
    \bigl(x_1^{2}+x_2^{2}\bigr)
    \exp\left[-\frac{r\bigl(x_1^{2}+x_2^{2}\bigr)}{2D_0}\right].
    \label{eq:SigU}
\end{align}
For the iTUR in the main text, we additionally require 
$\langle Q\rangle_\mu$ and $\E^\mu[\overline{J}_t]$.
For given system parameters we again solve numerically
\begin{align}
    \langle Q\rangle_\mu
= \frac{2r}{\pi^{3}h^{4}}
    \int_{z_1-h}^{z_1+h} d x_1
    \int_{z_2-\sqrt{h^{2}-(x_1-z_1)^{2}}}^{z_2+\sqrt{h^{2}-(x_1-z_1)^{2}}}
    d x_2\left[1-\frac{(x_1-z_1)^{2}+(x_2-z_2)^{2}}{h^{2}}\right]^{2}
    \exp\left[-\frac{r\bigl(x_1^{2}+x_2^{2}\bigr)}{2D_0}\right],
    \label{eq:Qmean}
\end{align}
as well as
\begin{align}
    \E^\mu[\overline{J}_t]
    = \frac{\Omega r}{\pi^{2}h^{2}D_0}
    \int_{z_1-h}^{z_1+h} d x_1
    \int_{z_2-\sqrt{h^{2}-(x_1-z_1)^{2}}}^{z_2+\sqrt{h^{2}-(x_1-z_1)^{2}}} d x_2
    \left[1-\frac{(x_1-z_1)^{2}+(x_2-z_2)^{2}}{h^{2}}\right]
    \bigl(n_1 x_2 - n_2 x_1\bigr)
    \exp\left[-\frac{r\bigl(x_1^{2}+x_2^{2}\bigr)}{2D_0}\right].
    \label{eq:meanJ}
\end{align}

\subsection{Simulation parameters}
Throughout, we generate trajectories of Eq.~\eqref{eq:OUP}
using a simple Euler-Maruyama scheme with time step $dt=10^{-3}$.
All trajectories are initialized from
their stationary state $\nu=\mu$ hence $N_\nu=1$, i.e., 
they are sampled from $\ps(x)$.
Moreover, we consider Eq.~\eqref{eq:OUP} with
$r=\lambda_\mathrm{gap}=1$ and $D_0=1$.
For dynamics with $\Omega=0$ and $\Omega=4$
we simulate $10^4$ independent trajectoires each of length $t=500$.
For $\Omega=25$ we use $2\times10^4$ trajectories each of duration $t=100$.
Generalized currents are computed from trajectories
using a discretized Stratonovich integral using the midpoint convention
\begin{align}
    \overline{J}_t
    \simeq \frac{1}{t}\sum_{k=0}^{t/ dt-1}
    U\left(\frac{X_{t_k}+X_{t_{k+1}}}{2}\right)\cdot
    \bigl(X_{t_{k+1}}-X_{t_k}\bigr).
    \label{eq:current_sum}
\end{align}

\begin{itemize}

    \item For Fig.~1c we consider Eq.~\eqref{eq:OUP}
    with $\Omega=4$ (blue) and $\Omega=0$ (red)
    and show trajectories of duration $t=7.5$.
    The steady-state current $\js$ (yellow) 
    is given by Eq.~\eqref{eq:js_OUP}.
    For the window function $V_z^h$ we use Eq.~\eqref{eq:EpaKernel}
    with $h=0.5$ and $z=(2,0)^\top$.
    We further use $\hat{n}=\hat{e}_1 =(1,0)^\top$ (green)
    and $\hat{n}=-\hat{e}_2 =(0,-1)^\top$ (magenta).
    In Fig.~1d we show corresponding empirical currents $\overline{J}_t$
    obtained via Eq.~\eqref{eq:current_sum} from trajectories of length $t=100$.
    Corresponding stationary means are computed numerically
    via Eq.~\eqref{eq:meanJ} using the respective parameters
    and read $\E^\mu[\overline{J}_t] = 0$ for $\Omega =0$
    and $\E^\mu[\overline{J}_t] = 0.172$ for $\Omega =4$.
    
    \item For Fig.~2 we consider Eq.~\eqref{eq:OUP}
    with $\Omega=4$ (Fig.~2a-d)
    and $\Omega=0$ (Fig.~2c)
    with $V_z^h$ from Eq.~\eqref{eq:EpaKernel}
    where we use
    $h=1$, $z=(1,0)^\top$, and $\hat{n}=-\hat{e}_2=(0,-1)^\top$.
    The parameters 
    $\|Q\|_{L^\infty(\mu)}$, $\Sigma^U$, and $\Sigma^U_\infty$ entering the bound are computed via Eqs.~\eqref{eq:supnorm_konschdanden} and~\eqref{eq:SigU}
    and read $\|Q\|_{L^\infty(\mu)}=0.405$, 
    $\Sigma^U=4.422$, and 
    $\Sigma^U_\infty=64$,
    respectively, for our parameters.
    
    \item For Fig.~3a 
    we consider Eq.~\eqref{eq:OUP} with $\Omega=25$
    and for $V_z^h$ from Eq.~\eqref{eq:EpaKernel}
    use $h=1.5$, $z=(0.5,0)^\top$, and $\hat{n}=-\hat{e}_2=(0,-1)^\top$.
    We compare it with the corresponding equilibrium bound
    which further requires the parameter $\|Q\|_{L^\infty(\mu)}=0.08$
    from Eq.~\eqref{eq:supnorm_konschdanden}
    for the values under consideration.

    \item For Fig.~3b 
    we consider Eq.~\eqref{eq:OUP} with $\Omega=4$
    and for $V_z^h$ from Eq.~\eqref{eq:EpaKernel}
    use $h=1$, $z=(1,0)^\top$, and $\hat{n}=-\hat{e}_2=(0,-1)^\top$.
    We empirically compute $\Var_{\mu}(\overline{J}_t)$
    from the sampled $\overline{J}_t$ as a function of time $t$
    and for the bounds use
    $\|Q\|_{L^\infty(\mu)}=0.405$, 
    $\Sigma^U=4.422$, and 
    $\Sigma^U_\infty=64$
    as before.
    Further with Eqs.~\eqref{eq:epr}, \eqref{eq:Qmean} and~\eqref{eq:meanJ}
    we compute
    $\Sigma=32$,
    $\langle Q\rangle_\mu=0.038$,
    and
    $\E^\mu[\overline{J}_t]=0.302$.

    \item For Fig.~3c,d 
    we consider Eq.~\eqref{eq:OUP} with $\Omega=4$
    and for $V_z^h$ from Eq.~\eqref{eq:EpaKernel}
    use $h=1$, $z=(1,0)^\top$, and $\hat{n}=-\hat{e}_2=(0,-1)^\top$.
    As before the parameters entering
    the uncertainty results for the given choice
    read  $\|Q\|_{L^\infty(\mu)}=0.405$, 
    $\Sigma^U=4.422$, and 
    $\Sigma^U_\infty=64$.
\end{itemize}

\subsection{Gaussian (central-limit) approximation}
In the main text we compare our results with a Gaussian (central-limit)
approximation. Hereby, we replace the distribution of
$\overline{J}_t$ with a normal distribution
$\mathcal{N}(\E^\mu[\overline{J}_t], \sigma_{J,\infty}^2/t)$
with stationary mean $\E^\mu[\overline{J}_t]$ and asymptotic variance 
$\sigma_{J,\infty}^2\equiv\lim_{t\to\infty} t\Var_\mu(\overline{J}_t)$.
The asymptotic variance is obtained from the long-time (saturated) value 
of the respective empirical $t\Var_\mu(\overline{J}_t)$.

\section{Proof strategy for the general concentration inequality}
Here we outline the basic ideas and relevant intermediate results
of the proof strategy 
used in \cite{Arxiv_Stratonovich} 
to derive the general concentration inequality for additive functionals of Stratonovich type.
As in the main text, throughout, $U$ is a bounded differentiable vector field 
and $\nu\ll\mu$ with $d\nu/d\mu\in L^{2}(\mu)$. 
Further, 
we consider $\overline{J}_t=t^{-1}\int_0^tU(X_s)\circ d X_s$
and use the local quantities $W =\vs\cdot U$
and $Q = U\cdot DU$.

\subsection{Cram\'er-Chernoff method and Feynman-Kac semigroup}
Our proof starts from Chernoff's inequality \cite{boucheron2013concentration},
which gives an upper bound on the tail probability in terms of the 
moment-generating function of the generalized current $\overline{J}_t$
via 
\begin{align}
    \mathbb{P}^\nu\bigl(\overline{J}_t-\E^\mu[\overline{J}_t]\ge a\bigr)
    \leq\inf_{k\ge0}\e^{-kta}
    \E^\nu\bigl[\e^{kt(\overline{J}_t-\E^\mu[\overline{J}_t])}\bigr]
    =\inf_{k\ge0}\e^{-kt(\E^\mu[\overline{J}_t]+a)} \E^\nu\bigl[\e^{kJ_t}\bigr],
    \label{eq:general_chernoff}
\end{align}
where $k$ is a tilting parameter
and we denote the accumulated current as $J_t \equiv t \overline{J}_t$.
Hence, to prove the required tail bound, we have to establish control 
over the moment-generating function 
$\E^\nu\bigl[\e^{k J_t}\bigr]$, which will be done in the next steps
that follow.
Our first result in \cite{Arxiv_Stratonovich}
is a bound of the moment-generating function based
on the Cauchy-Schwarz inequality, which reads
\begin{align}
  \E^\nu\bigl[\e^{kJ_t}\bigr]
  \leq
  N_\nu \bigl\|P_t^{k,J} \bigr\|_{L^2(\mu)},
\end{align}
where $N_\nu= \|d\nu/d\mu\|_{L^2(\mu)}$ is the same prefactor accounting 
for the initial condition as in the main text, with $N_\nu=1$ if and only if $\nu=\mu$.
Key to establishing this bound is the operator norm
$\|P_t^{k,J}\|_{L^2(\mu)}$ on $L^2(\mu)$
of the corresponding Feynman-Kac semigroup 
for current-type observables
\begin{align}
  \bigl(P_t^{k,J} f\bigr)(x)
  \equiv
  \E^x\left[
    f(X_t)\exp\left(k \int_0^t U(X_s)\circ d X_s\right)
  \right],
  \label{eq:fk_current}
\end{align}
which is generated by a tilted generator $\mathcal{L}^J_k$ \cite{dieball2023feynman}.
In a crucial next step, based on the Lumer-Phillips
theorem \cite{engel2000one}, we show that the operator norm of the Feynman-Kac semigroup 
is bounded according to 
\begin{align}
  \|P_t^{k,J}\|_{L^2(\mu)}
  \leq
  \e^{t\Lambda^J(k)}
  \implies
  \E^\nu\bigl[\e^{kJ_t}\bigr]
  \leq
  N_\nu
  \e^{t\Lambda^J(k)}.
  \label{eq:fk_bound}
\end{align}
In particular, 
we show that
$\Lambda^J(k)=\lambda_{\max}\bigl(\widetilde{\mathcal{L}}_k^J\bigr)$, i.e., it
corresponds to the largest eigenvalue of 
the \emph{symmetrized} tilted current operator 
$\widetilde{\mathcal{L}}^J_k\equiv\tfrac12(\mathcal{L}^J_k
+\mathcal{L}^{J\dagger}_k)$ [with $\dagger$ the adjoint in $L^{2}(\mu)$],
which we show takes the form
\begin{align}
    \widetilde{\mathcal{L}}_k^Jf
    = \mathcal{L}_S f + k(\vs\cdot U) f + k^2(U\cdot DU)f,
\end{align}
where $\mathcal{L}_S$ is the reversible generator of
the underlying Markov process from the main text.
Taken together, inserting
Eq.~\eqref{eq:fk_bound} back into Eq.~\eqref{eq:general_chernoff}
thus gives
\begin{align}
    \mathbb{P}^\nu\bigl(\overline{J}_t - \E^\mu[\overline{J}_t]\geq a\bigr)
    \leq
    \inf_{k\geq 0}
    N_\nu\e^{-t \bigl[ka+ k\E^\mu[\overline{J}_t] - \Lambda^J(k)\bigr]}
    = N_\nu\e^{-t \sup_{k\geq 0} \bigl[ka+ k\E^\mu[\overline{J}_t] - \Lambda^J(k)\bigr]}.
    \label{eq:chernoff_unopt}
\end{align}
To carry out the optimization over $k$
explicitly in the last step,
we subsequently establish an upper bound on $\Lambda^J(k)$
based on the Poincar\'e inequality via two different strategies (see \cite{Arxiv_Stratonovich}
for details).

\subsubsection{Approach 1 to bound $\Lambda^J(k)$}
In the first approach we prove that, for all
$0 \leq k < \lambda_\mathrm{gap}/\|W-\langle W\rangle_\mu\|_{L^\infty(\mu)}$,
the following bound holds \cite{Arxiv_Stratonovich},
\begin{align}
    \Lambda^J(k)
    \leq
    k\E^\mu[\overline{J}_t]
+ \frac{k^2 \bigl(\mathrm{Var}_\mu(W) + \lambda_\mathrm{gap}\|Q\|_{L^\infty(\mu)}\bigr)}{\lambda_\mathrm{gap} - k \|W - \langle W\rangle_\mu\|_{L^\infty(\mu)}}
= k\E^\mu[\overline{J}_t]
    + \frac{k^2\widetilde{\mathcal{C}}}{2(1-k\widetilde{\mathcal{D}})},
    \label{eq:upper_bound1}
\end{align}
where the second equality follows by rearranging 
and identifying the constants
$\widetilde{\mathcal{C}}$ and $\widetilde{\mathcal{D}}$ of
Eq.~(A2), i.e., 
\begin{align}
    \widetilde{\mathcal{C}}
    \equiv 2 \| Q \|_{L^\infty(\mu)} + \frac{2\Var_\mu(W)}{\lambda_\mathrm{gap}},
    \quad
    \widetilde{\mathcal{D}}
    \equiv \frac{\| W-\langle W\rangle_\mu \|_{L^\infty(\mu)}}{\lambda_\mathrm{gap}}.
    \label{eq:aap_constants}
\end{align}
Moreover, with these constants, the admissible range now reads
$0\leq k<1/\widetilde{\mathcal{D}}$.
Subsequently, 
inserting the upper bound~\eqref{eq:upper_bound1} 
into Eq.~\eqref{eq:chernoff_unopt} (note that the two terms
$k\E^\mu[\overline{J}_t]$ cancel)
and carrying out the optimization in
the parameter $k$ gives 
\begin{align}
    \sup_{0\leq k<1/\widetilde{\mathcal{D}}}
    \left[ka-\frac{k^2\widetilde{\mathcal{C}}}{2(1-k\widetilde{\mathcal{D}})}\right]
    =
    \frac{\widetilde{\mathcal{C}}}{\widetilde{\mathcal{D}}^{2}}
    h\left(\frac{\widetilde{\mathcal{D}}a}{\widetilde{\mathcal{C}}}\right),
\end{align}
where $h(u)=1+u-\sqrt{1+2u}$.
This then proves the general concentration inequality of Eq.~(A1).

\subsubsection{Approach 2 to bound $\Lambda^J(k)$}
In the second approach, we derive
a different bound on $\Lambda^J$.
To this end, let
$\bar W = W - \langle W\rangle_\mu$ and
$\bar Q = Q - \langle Q\rangle_\mu$.
Further denote $k^\ast_{\max}$ as
the positive root of
$k(\|\bar W\|_{L^\infty(\mu)} + k\|\bar Q\|_{L^\infty(\mu)}) = \lambda_\mathrm{gap}$, i.e.,
\begin{align}
k^\ast_{\max}
=
\begin{cases}
\dfrac{-\|\bar W\|_{L^\infty(\mu)} + \sqrt{\|\bar W\|_{L^\infty(\mu)}^2 + 4\lambda_\mathrm{gap}\|\bar Q\|_{L^\infty(\mu)}}}{2\|\bar Q\|_{L^\infty(\mu)}}
& \|\bar Q\|_{L^\infty(\mu)} > 0, \\[6pt]
\dfrac{\lambda_\mathrm{gap}}{\|\bar W\|_{L^\infty(\mu)}}
& \|\bar Q\|_{L^\infty(\mu)} = 0.
\end{cases}
\label{eq:kmaxx}
\end{align}
In \cite{Arxiv_Stratonovich} 
we prove that for all  $0\leq k<k^\ast_{\max}$
the following bound holds, 
\begin{align}
    \Lambda^J(k)
    \leq
    k\E^\mu[\overline{J}_t]
    + \frac{k^{2}\widetilde{\mathcal{C}}_2}{2(1-k\widetilde{\mathcal{D}}_2)},
    \label{eq:lambda_bound2}
\end{align}
where the new parameters read
\begin{align}
    \widetilde{\mathcal{C}}_2
    \equiv 2\langle Q\rangle_\mu
    + \frac{2}{\lambda_\mathrm{gap}}
      \left(\sqrt{\Var_\mu(W)}
      + k^\ast_{\max}\sqrt{\Var_\mu(Q)}\right)^{2},
    \qquad
    \widetilde{\mathcal{D}}_2\equiv\frac{1}{k^\ast_{\max}},
\label{eq:upper_bound2}
\end{align}
and the admissible range becomes $0\leq k<1/\widetilde{\mathcal{D}}_2$.
Since Eq.~\eqref{eq:lambda_bound2} has 
the same (sub-gamma) structure as Eq.~\eqref{eq:upper_bound1},
and the mean terms again cancel,
the optimization in $k$
gives analogously
\begin{align}
    \sup_{0\leq k<1/\widetilde{\mathcal{D}}_2}
    \left[ka-\frac{k^{2}\widetilde{\mathcal{C}}_2}{2(1-k\widetilde{\mathcal{D}}_2)}\right]
    =\frac{\widetilde{\mathcal{C}}_2}{\widetilde{\mathcal{D}}_2^{2}}
    h\left(\frac{\widetilde{\mathcal{D}}_2a}{\widetilde{\mathcal{C}}_2}\right).
\end{align}
The general bound~(A1)
thus also holds with the constants $\widetilde{\mathcal{C}}_2$ and 
$\widetilde{\mathcal{D}}_2$.
Notably, neither of the two approaches is uniformly sharper as
we discuss in \cite{Arxiv_Stratonovich}.
The main takeaway is that 
Approach 2 sharpens via
$\|Q\|_{L^\infty(\mu)}\to\langle Q\rangle_\mu$,
but increases the exponential tail regime
$\widetilde{\mathcal{D}}_2\geq\widetilde{\mathcal{D}}$.

\subsection{Extension to the sample mean}
Our proof strategy, based on the Cram\'er-Chernoff method \cite{boucheron2013concentration},
extends straightforwardly to the sample mean 
$\widehat{J}_{n,t}=n^{-1}\sum_{i=1}^{n}\overline{J}^{(i)}_t$ over $n$
statistically independent trajectories of length $t$.
Hereby, each trajectory is started from 
the same initial law $\nu$.
The underlying reason is that the method solely rests on the moment-generating function, 
which factorizes for independent realizations.
Concretely, using either of the two established bounds on $\Lambda^J(k)$, the (centered) moment-generating function of a single time-average has the sub-gamma bound
\begin{align}
    \E^\nu\bigl[\e^{kt(\overline{J}_t-\E^\mu[\overline{J}_t])}\bigr]
    \leq N_\nu\exp\left[\frac{t\widetilde{\mathcal{C}}k^{2}}
    {2(1-\widetilde{\mathcal{D}}k)}\right],
    \label{eq:oifach}
\end{align}
or alternatively with the parameters $\widetilde{\mathcal{C}}_2$ and $\widetilde{\mathcal{D}}_2$
from the second approach.
In analogy to the single-trajectory setting, we can introduce the corresponding moment-generating 
function of the sample mean
with (tilting)
parameter $s\equiv knt$, $0 < s < nt/\widetilde{\mathcal{D}}$,
\begin{align}
        \E^\nu\bigl[\e^{s(\widehat{J}_{n,t}-\E^\mu[\overline{J}_t])}\bigr]
    =\prod_{i=1}^{n}
    \E^\nu\bigl[\e^{(s/n)(\overline{J}^{(i)}_t-\E^\mu[\overline{J}_t])}\bigr]
    \le N^{n}_\nu\exp\left\{
    \frac{(\widetilde{\mathcal{C}}/nt)s^{2}}
    {2\bigl[1-(\widetilde{\mathcal{D}}/nt)s\bigr]}\right\},
\end{align}
where the first equality follows from independence of the individual realizations,
and the inequality follows from applying Eq.~\eqref{eq:oifach} 
with $kt=s/n$ for each factor.
Consequently, the sample mean $\widehat{J}_{n,t}$ is sub-gamma
with a variance proxy of $\widetilde{\mathcal{C}}/nt$
and exponential scale parameter $\widetilde{\mathcal{D}}/nt$,
while the prefactor now reads $N_\nu^n$.
Carrying out the same optimization as before yields,
for all $n\geq 1$, $t > 0$, $a\geq 0$,
\begin{align}
    \mathbb{P}^\nu\bigl(\widehat{J}_{n,t}-\E^\mu[\overline{J}_t]\geq a\bigr)
    \leq N^{n}_\nu\exp\bigl[-nt\mathcal{I}^J(a)\bigr],
\end{align}
where $\mathcal{I}^J(a)$ is the same function as for the single-trajectory case, i.e., 
results for a single realization can be applied to the sample-mean setting
with the substitutions $t\to nt$ and $N_\nu\to N_\nu^n$.
Notably, whereas the exponent grows linear in $n$ and also in $t$,  
the prefactor grows exponentially in $n$ alone.
This again highlights that the initial condition cannot be averaged away by increasing the sample size as we already discussed in App.~D.

\section{Proof of thermodynamic concentration inequality}
The thermodynamic concentration inequality of Eq.~(9)
in the main text follows from upper bounding the parameters of the general
bound of \cite{Arxiv_Stratonovich} (obtained via the first approach) in terms of
physical quantities, i.e., our goal is to bound
\begin{align}
    \widetilde{\mathcal{C}}
    \equiv 2\|Q\|_{L^\infty(\mu)} + \frac{2\Var_\mu(W)}{\lambda_\mathrm{gap}},
    \qquad
    \widetilde{\mathcal{D}}
    \equiv \frac{\|W-\langle W\rangle_\mu\|_{L^\infty(\mu)}}{\lambda_\mathrm{gap}}.
    \label{eq:general_konschdande}
\end{align}
The proof relies on bounding the term $W=\vs \cdot U$
in terms of parameters related to the steady-state entropy production rate.
Namely, $W$ couples the observable $U$ 
to the nonequilibrium character of the dynamics $\vs=\js /\ps$,
which vanishes identically at detailed balance and is nonzero
out of equilibrium.
Note further that 
both $W=W\mathbb{1}_{\mathrm{supp}(U)}$ and $Q=Q\mathbb{1}_{\mathrm{supp}(U)}$ 
since $W$ and $Q$ vanish by construction
outside $\mathrm{supp}(U)$, i.e.,
the indicator function $\mathbb{1}_{\mathrm{supp}(U)}$
can be inserted freely and renders the resulting constants local to the support of $U$.
Writing $W=\vs \cdot U = (D^{-1/2} \vs) \cdot (D^{1/2}U)$ and using the Cauchy-Schwarz inequality gives
\begin{align}
    W^2\leq |D^{-1/2} \vs|^2  |D^{1/2}U|^2 = (\vs \cdot D^{-1} \vs)(U\cdot D U)
    = \Sigmaloc \mathbb{1}_{\mathrm{supp}(U)} Q.
    \label{eq:Wsquare_bound1}
\end{align}
or equivalently
\begin{align}
    |W(x)|\leq\sqrt{\Sigma_\mathrm{loc}(x)\mathbb{1}_{\mathrm{supp}(U)}(x)Q(x)},
    \label{eq:W_bound1}
\end{align}
where $\Sigmaloc =\vs\cdot D^{-1}\vs$ as before.
We first bound the variance term by taking the average over $W^2$,
which gives two different choices for possible upper bounds
\begin{align}
    \langle W^2\rangle_\mu
    &\leq 
    \langle \Sigma_\mathrm{loc}\mathbb{1}_{\mathrm{supp}(U)}Q\rangle_\mu
    \leq 
    \langle Q\rangle_\mu
    \|\Sigma_\mathrm{loc}\mathbb{1}_{\mathrm{supp}(U)} \|_{L^\infty(\mu)}
    = 
    \Sigma^U_\infty \langle Q\rangle_\mu,
    \\
    \langle W^2\rangle_\mu
    &\leq 
    \langle \Sigma_\mathrm{loc}\mathbb{1}_{\mathrm{supp}(U)}Q\rangle_\mu
    \leq 
    \| Q \|_{L^\infty(\mu)} 
    \langle \Sigma_\mathrm{loc}\mathbb{1}_{\mathrm{supp}(U)} \rangle_\mu
    =
    \Sigma^U
    \| Q \|_{L^\infty(\mu)},
\end{align}
where the first inequalities
follow from Eq.~\eqref{eq:W_bound1},
and the second inequalities follow by factoring out $\Sigmaloc$ or $Q$
via their respective sup-norm.
Hence, introducing $\langle\Sigma^UQ\rangle^{\infty}_{\wedge\vee}
\equiv\min\bigl\{\Sigma^U_\infty\langle Q\rangle_\mu,\;
\Sigma^U\|Q\|_{L^\infty(\mu)}\bigr\}$
to always select the best bound,
we have
\begin{align}
   \langle W^2\rangle_\mu \leq \langle\Sigma^UQ\rangle^{\infty}_{\wedge\vee},
   \label{eq:w2_bound}
\end{align}
and can therefore bound the variance according to 
\begin{align}
    \Var_\mu(W) = \langle W^2 \rangle_\mu - \langle W \rangle_\mu^2
    \leq
    \langle\Sigma^UQ\rangle^{\infty}_{\wedge\vee} - \E^\mu[\overline{J}_t]^2
    \leq 
    \langle\Sigma^UQ\rangle^{\infty}_{\wedge\vee} 
    \leq
    \Sigma^U\|Q\|_{L^\infty(\mu)},
    \label{eq:var_bound1}
\end{align}
where in the first inequality we use 
Eq.~\eqref{eq:w2_bound} to bound $\langle W^2 \rangle_\mu$
and that $\langle W\rangle_\mu = \E^\mu[\overline{J}_t]$.
In the second inequality we drop the mean contribution since $\E^\mu[\overline{J}_t]^2\geq 0$, 
and in 
the third inequality we only keep the second entry of the minimum.
Inserting Eq.~\eqref{eq:var_bound1} into $\widetilde{\mathcal{C}}$
of Eq.~\eqref{eq:general_konschdande}
we obtain
\begin{align}
    \widetilde{\mathcal{C}}
    \leq
    \mathcal{C}'
    =2\left(\|Q\|_{L^\infty(\mu)}
    +\frac{\langle\Sigma^UQ\rangle^{\infty}_{\wedge\vee}
    -\E^\mu[\overline{J}_t]^{2}}{\lambda_\mathrm{gap}}\right)
    \leq
    \mathcal{C}
    =2\|Q\|_{L^\infty(\mu)}
    \left(1+\frac{\Sigma^U}{\lambda_\mathrm{gap}}\right).
\end{align}
In the last bound we recover $\mathcal{C}$ from the main text, which trades
sharpness with a simpler form.
Here, the dissipation enters through the single factor $(1+\Sigma^U/\lambda_\mathrm{gap})$ that multiplies the corresponding equilibrium value
$\mathcal{C}^{\mathrm{eq}}=2\|Q\|_{L^\infty(\mu)}$.

It remains to bound the $\widetilde{\mathcal{D}}$ term.
We start with $\|W-\langle W\rangle_\mu \|_{L^\infty(\mu)}$ and
use the triangle inequality to obtain
\begin{align}
    \|W-\langle W\rangle_\mu \|_{L^\infty(\mu)}
    \leq 
    \| W \|_{L^\infty(\mu)}
    + |\langle W\rangle_\mu|.
\end{align}
Using Eq.~\eqref{eq:W_bound1} we can bound the first term via
\begin{align}
    \| W \|_{L^\infty(\mu)}
    \leq \| \sqrt{\Sigma_\mathrm{loc}\mathbb{1}_{\mathrm{supp}(U)}Q}\|_{L^\infty(\mu)}
    \leq 
    \sqrt{\Sigma_\infty^U \|Q \|_{L^\infty(\mu)}}.
\end{align}
Next, we bound the second term using Eq.~\eqref{eq:W_bound1}
together with Jensen's inequality 
and the Cauchy-Schwarz inequality by
\begin{align}
    |\langle W\rangle_\mu|
    \leq 
    \langle|W|\rangle_\mu
    \leq 
    \left\langle \sqrt{\Sigma_\mathrm{loc}\mathbb{1}_{\mathrm{supp}(U)}Q}\right\rangle_\mu
    \leq 
    \sqrt{\langle\Sigma_\mathrm{loc}\mathbb{1}_{\mathrm{supp}(U)}\rangle_\mu 
    \langle Q\rangle_\mu}
    = \sqrt{\Sigma^U  \langle Q\rangle_\mu}.
\end{align}
Thus, together we arrive at
\begin{align}
    \|W-\langle W\rangle_\mu \|_{L^\infty(\mu)}
    \leq 
    \sqrt{\Sigma_\infty^U \|Q \|_{L^\infty(\mu)}} 
    + \sqrt{\Sigma^U  \langle Q\rangle_\mu}
    \leq
    2 \sqrt{\Sigma_\infty^U \|Q \|_{L^\infty(\mu)}} ,
    \label{eq:w_inf_bound}
\end{align}
where the last inequality follows
from $\Sigma^U\leq\Sigma^U_\infty$ and $\langle Q\rangle_\mu\leq\|Q\|_{L^\infty(\mu)}$.
Together, the corresponding parameter is hence bounded by
\begin{align}
    \widetilde{\mathcal{D}}
    \leq\mathcal{D}'
    =\frac{\sqrt{\Sigma^U_\infty\|Q\|_{L^\infty(\mu)}}
    +\sqrt{\Sigma^U\langle Q\rangle_\mu}}{\lambda_\mathrm{gap}}
    \leq\mathcal{D}
    =\frac{2\sqrt{\Sigma^U_\infty\|Q\|_{L^\infty(\mu)}}}{\lambda_\mathrm{gap}},
\end{align}
which completes the proof of the thermodynamic concentration inequality of the main text,
and the additional version of App.~B.

\subsection{Additional thermodynamic concentration inequality based on Approach 2}
Bounding the general constants via thermodynamic quantities is not restricted 
to the general bound obtained from Approach 1.
Indeed, the same idea can be applied to the parameters $\widetilde{\mathcal{C}}_2$
and $\widetilde{\mathcal{D}}_2$ of the second approach.
Again, only the $W$ terms
in Eqs.~\eqref{eq:upper_bound2} and~\eqref{eq:kmaxx}
couple to the nonequilibrium character of the dynamics, i.e., 
we have to bound $\Var_\mu(W)$ and $\|\bar W\|_{L^\infty(\mu)}$.
In particular, it suffices to use the bounds~\eqref{eq:var_bound1}
and~\eqref{eq:w_inf_bound} derived above
\begin{align}
    \Var_\mu(W)
    &\leq\langle\Sigma^UQ\rangle^{\infty}_{\wedge\vee}
    -\E^\mu[\overline{J}_t]^2, 
    \\
    \|\bar W\|_{L^\infty(\mu)}
    &=\|W - \langle W\rangle_\mu\|_{L^\infty(\mu)}
    \leq2\sqrt{\Sigma^U_\infty\|Q\|_{L^\infty(\mu)}}.
    \label{eq:w_bound2}
\end{align}
Since $k^\ast_{\max}$ is decreasing in 
$\|\bar W\|_{L^\infty(\mu)}$
using Eq.~\eqref{eq:w_bound2} yields a new admissible range 
$k^\ast_{\mathrm{th}}\leq k^\ast_{\max}$, where
\begin{align}
    k^\ast_{\mathrm{th}}
    \equiv
    \begin{cases}
    \dfrac{\sqrt{\Sigma^U_\infty\|Q\|_{L^\infty(\mu)}
    +\lambda_\mathrm{gap}\|\bar Q\|_{L^\infty(\mu)}}
    -\sqrt{\Sigma^U_\infty\|Q\|_{L^\infty(\mu)}}}{\|\bar Q\|_{L^\infty(\mu)}}
    & \|\bar Q\|_{L^\infty(\mu)}>0,
    \\[10pt]
    \dfrac{\lambda_\mathrm{gap}}{2\sqrt{\Sigma^U_\infty\|Q\|_{L^\infty(\mu)}}}
    & \|\bar Q\|_{L^\infty(\mu)}=0,
    \end{cases}
\end{align}
where $k^\ast_{\mathrm{th}}$ is the positive root of
$k\bigl(2\sqrt{\Sigma^U_\infty\|Q\|_{L^\infty(\mu)}}
+k\|\bar Q\|_{L^\infty(\mu)}\bigr)=\lambda_\mathrm{gap}$.
The corresponding thermodynamic constants then read
\begin{align}
    \mathcal{C}_2
    \equiv2\langle Q\rangle_\mu
    +\frac{2}{\lambda_\mathrm{gap}}
    \left(\sqrt{\langle\Sigma^UQ\rangle^{\infty}_{\wedge\vee}
    -\E^\mu[\overline{J}_t]^2}
    +k^\ast_{\mathrm{th}}\sqrt{\Var_\mu(Q)}\right)^{2},
    \qquad
    \mathcal{D}_2\equiv\frac{1}{k^\ast_{\mathrm{th}}},
\end{align}
and Eq.~(9)
holds with $\mathcal{C}_2$ and $\mathcal{D}_2$ as well
since 
carrying out the optimization on the smaller interval $0\leq k<k^\ast_{\mathrm{th}}$
leads to the same sub-gamma form of the bound on $\Lambda^J(k)$.
Finally, we remark that using 
the sharper bound 
$\|\bar W\|_{L^\infty(\mu)}\leq\sqrt{\Sigma^U_\infty\|Q\|_{L^\infty(\mu)}}
+\sqrt{\Sigma^U\langle Q\rangle_\mu}$ instead of Eq.~\eqref{eq:w_bound2}
yields valid bounds with sharper, but more involved, constants.

\section{Detecting broken detailed balance}
As we outline in App.~F,
a statistically significant violation of 
the equilibrium form of the 
thermodynamic concentration inequality 
can be used to detect irreversibility.
However, since the required empirical tail probability
is a random quantity itself, a rigorous test to check whether
the equilibrium bound is violated hinges on 
an appropriate quantification of the accompanying uncertainty of the inferred estimate.
In line with the spirit of the paper, we will achieve this
in a nonasymptotic and model-free manner (i.e., without invoking a Gaussian or asymptotic approximation)
via Hoeffding's inequality \cite{boucheron2013concentration}.

To this end, let us assume we are given $n$ independent trajectories $(X_s^{(i)})_{0\leq s\leq t}$ each of fixed length $t$, that 
are initialized from the same initial measure $\nu$.
For the procedure of App.~F, we choose some admissible 
$U$ and fix a threshold $a >0$.
The relevant random variable is then the
bounded variable $Z_i\in[0,1]$ defined by
\begin{align}
    Z_i \equiv 
    \begin{cases}
        1\quad \text{for}\quad\overline{J}_t^{(i)}\geq a,
        \\
        0\quad \text{otherwise},
    \end{cases}
\end{align}
such that 
$P_a^\nu\equiv\mathbb{P}^\nu(\overline{J}_t\geq a)$
is estimated via
the empirical tail probability given by
\begin{align}
    \hat{P}_a^\nu\equiv \frac{1}{n}\sum_{i=1}^n Z_i = \frac{n_a}{n},
    \label{eq:emp_prob_sample_mean}
\end{align}
where $n_a$ is the number of trajectories for which $\overline{J}_t^{(i)}\geq a$.
Notably, $Z_i$ has mean $\E[Z_i]=P_a^\nu$ and therefore
$\E[\hat{P}_a^\nu] = P_a^\nu$, i.e., $\hat{P}_a^\nu$ is an unbiased estimator.
Consequently, applying Hoeffding's inequality to the sample mean~\eqref{eq:emp_prob_sample_mean}
gives
\begin{align}
       \mathbb{P}(\hat{P}_a^\nu -P_a^\nu \geq \varepsilon)\leq \e^{-2n\varepsilon^2}. 
\end{align}
Setting the right-hand side 
equal to $\alpha$, with $\alpha\in(0,1)$,
and solving for $\varepsilon$ results in 
\begin{align}
    \varepsilon= \sqrt{\frac{\log(1/\alpha)}{2n}},
\end{align}
such that with a probability of at least $1-\alpha$ under $\mathbb{P}^\nu$
it holds that
\begin{align}
    P_a^\nu \geq \hat{P}_a^\nu- \sqrt{\frac{\log(1/\alpha)}{2n}}.
    \label{eq:uq_detect}
\end{align}
As discussed in the main text and App.~F,
under detailed balance
$\mathcal{D}^\mathrm{eq}=0$, 
$\mathcal{C}^\mathrm{eq}= 2 \| Q \|_{L^\infty(\mu)}$,
and $\E^\mu[\overline{J}_t]=0$, such that 
the thermodynamic concentration inequality
reduces to the purely sub-Gaussian form
\begin{align}
    P_a^\nu \leq N_\nu \exp\bigl[ -t \mathcal{I}_\mathrm{eq}^J(a) \bigr],
    \qquad 
    \mathcal{I}_\mathrm{eq}^J(a) = \frac{a^2}{4 \| Q \|_{L^\infty(\mu)}}.
    \label{eq:uq_db_bound}
\end{align}
By taking finite-sample effects
of the empirical tail probability into account via Eq.~\eqref{eq:uq_detect},
we can thus conclusively say
that whenever we detect (see Fig.~3a of the main text)
\begin{align}
    \hat{P}^\nu_a-\sqrt{\frac{\log(1/\alpha)}{2n}}
    >
    N_\nu\exp\left[-\frac{ta^{2}}{4\|Q\|_{L^\infty(\mu)}}
    \right],
    \label{eq:db_violation}
\end{align}
the fluctuations of $\overline{J}_t$
are \emph{inconsistent} with detailed balance [Eq.~\eqref{eq:uq_db_bound}]
and we can conclusively rule out equilibrium 
of the underlying dynamics with a confidence of at least $1-\alpha$.
Note further that the test only requires
$\|Q\|_{L^\infty(\mu)}=\|U\cdot DU\|_{L^\infty(\mu)}$
where $U$ is chosen and $D$ is accessible from 
short-time behavior of the mean-squared displacement, i.e., 
neither knowledge of the underlying driving nor dissipation rate, nor spectral gap are 
required to apply the test.
Lastly, we remark that by construction our test is sufficient but not necessary 
for detecting irreversibility.
While satisfying Eq.~\eqref{eq:db_violation}
does detect broken detailed balance, a failure to satisfy it
does \emph{not} certify detailed balance, since
the underlying bound~\eqref{eq:uq_db_bound}
is not tight and the chosen $U$ may be blind to the current.

\section{Proof of dissipation concentration inequality}
Here we prove the dissipation concentration inequality corresponding
to Eq.~(11) in the main text.
Recall that for the dissipation concentration inequality
we consider a smooth window function 
$0\leq K^h_z\leq1$ localized at $z$ with scale $h$
and compact support $\mathrm{supp}(K^h_z)$.
Subsequently, we consider the vector field 
$U\equiv U_\Sigma = D^{-1} \vs K_z^h$, with $\mathrm{supp}(U_\Sigma)=\mathrm{supp}(K^h_z)$,
such that the associated Stratonovich 
functional becomes
\begin{align}
    \Sigmaline^K\equiv\overline{J}_t(U_\Sigma)
    =\frac{1}{t}\int_0^t\bigl[D^{-1}\vs K^h_z\bigr](X_s)\circ d X_s.
\end{align}
This choice of a generalized current corresponds to a trajectory estimator of the
coarse-grained regional entropy production rate 
$\E^\mu[\Sigmaline^K]=\langle\Sigma_\mathrm{loc}K^h_z\rangle_\mu\equiv\Sigma^K\leq\Sigma^U$.
The last inequality follows 
from $K^h_z\leq\mathbb{1}_{\mathrm{supp}(U_\Sigma)}$
and the limit $K^h_z\to\mathbb{1}_{\mathrm{supp}(U_\Sigma)}$
recovers $\Sigma^K\to\Sigma^U$.
For this particular choice of $U_\Sigma$ the corresponding
local parameters now read
\begin{align}
    W_\Sigma
    &=\vs\cdot U_\Sigma
    =(\vs\cdot D^{-1}\vs)K^h_z
    =\Sigma_\mathrm{loc}K^h_z,
    \label{eq:dissi_parameters1}
    \\
    Q_\Sigma
    &=U_\Sigma\cdot DU_\Sigma
    =(\vs\cdot D^{-1}\vs)(K^h_z)^{2}
    =\Sigma_\mathrm{loc}(K^h_z)^{2}.
    \label{eq:dissi_parameters}
\end{align}
Note that $W^2_\Sigma = \Sigmaloc Q_\Sigma$, i.e., Eq.~\eqref{eq:Wsquare_bound1}
holds with equality.

To prove the dissipation concentration inequality we start with the general concentration
inequality applied to the specific $U_\Sigma$
and the respective constants 
then read
\begin{align}
    \widetilde{\mathcal{C}}_\Sigma
    =2\|Q_\Sigma\|_{L^\infty(\mu)}+\frac{2\Var_\mu(W_\Sigma)}{\lambda_\mathrm{gap}},
    \qquad
    \widetilde{\mathcal{D}}_\Sigma =
    \frac{\|W_\Sigma-\langle W_\Sigma\rangle_\mu\|_{L^\infty(\mu)}}{\lambda_\mathrm{gap}}.
\end{align}
We start with deriving an upper bound on $\widetilde{\mathcal{C}}_\Sigma$.
For the first term it holds that
\begin{align}
    \|Q_\Sigma\|_{L^\infty(\mu)}
    =\bigl\|\Sigma_\mathrm{loc}(K^h_z)^2\bigr\|_{L^\infty(\mu)}
    \leq\bigl\|\Sigma_\mathrm{loc}\mathbb{1}_{\mathrm{supp}(U_\Sigma)}\bigr\|_{L^\infty(\mu)}
    =\Sigma^U_\infty,
    \label{eq:sup_Q_Sigma_bound}
\end{align}
where the inequality follows from  $(K_z^h)^2\leq \mathbb{1}_{\mathrm{supp}(U_\Sigma)}$
since $0\leq K^h_z \leq1$.
For the variance term we further have
\begin{align}
    \Var_\mu(W_\Sigma) =
    \langle(\Sigma_\mathrm{loc}K^h_z)^{2}\rangle_\mu - (\langle\Sigma_\mathrm{loc}K^h_z\rangle_\mu)^2
    \leq \Sigma_\infty^U\langle\Sigma_\mathrm{loc}K^h_z\rangle_\mu - (\langle\Sigma_\mathrm{loc}K^h_z\rangle_\mu)^2
    = \Sigma^U_\infty\Sigma^K-(\Sigma^K)^2,
\end{align}
where in the first inequality we pull out one factor $\Sigmaloc$ and bound it via
the supremum $\Sigma_\infty^U$ within the supported region,
and use that $(K_z^h)^2\leq K_z^h$.
Therefore, we have
\begin{align}
     \Var_\mu(W_\Sigma) \leq
     \Sigma^U_\infty\Sigma^K-(\Sigma^K)^2
     \leq
     \Sigma^U_\infty\Sigma^K
     \leq \Sigma^U_\infty\Sigma^U,
\end{align}
where the second inequality follows by 
dropping $(\langle\Sigma_\mathrm{loc}K^h_z\rangle_\mu)^2\geq0$,
and the third inequality follows from $\Sigma^K\leq\Sigma^U$.
Consequently, together with Eq.~\eqref{eq:sup_Q_Sigma_bound}, we have
\begin{align}
    \widetilde{\mathcal{C}}_\Sigma
    =2\|Q_\Sigma\|_{L^\infty(\mu)}+\frac{2\Var_\mu(W_\Sigma)}{\lambda_\mathrm{gap}}
    \leq 
    2\Sigma^U_\infty\left(1+\frac{\Sigma^K}{\lambda_\mathrm{gap}}\right)
    \leq   
    2\Sigma^U_\infty\left(1+\frac{\Sigma^U}{\lambda_\mathrm{gap}}\right)
    =\mathcal{C}_\Sigma.
\end{align}
It remains to bound the $\widetilde{\mathcal{D}}$ term.
Using Eq.~\eqref{eq:dissi_parameters},
we first introduce
\begin{align}
    \bar\Sigma^K_\infty
    \equiv\bigl\|\Sigma_\mathrm{loc}K^h_z-\Sigma^K\bigr\|_{L^\infty(\mu)}
    =
    \|W_\Sigma-\langle W_\Sigma\rangle_\mu\|_{L^\infty(\mu)}.
\end{align}
Since $\Sigmaloc \geq 0$ and $0\leq K_z^h \leq 1$,
we have $\Sigmaloc K_z^h \in[0,\Sigma_\infty^U]$
and consequently also $\Sigma^K=\langle \Sigmaloc K_z^h\rangle_\mu\in[0,\Sigma_\infty^U]$.
Therefore, we have
\begin{align}
    -\Sigma^K \leq 
    \Sigmaloc K_z^h - \Sigma^K \leq \Sigma^U_\infty -\Sigma^K.
\end{align}
Since for a generic $f\in[-A,B]$, $A, B \geq0$,
it holds that $\| f\|_{L^\infty(\mu)}\leq \max\{A,B\}$,
we have
\begin{align}
    \bar\Sigma^K_\infty =
    \|\Sigmaloc K_z^h - \Sigma^K\|_{L^\infty(\mu)}
    \leq
    \max\{\Sigma^K, \Sigma_\infty^U-\Sigma^K \}
    \leq \Sigma_\infty^U,
\end{align}
where the last inequality follows since $\Sigma^K\in[0,\Sigma^U_\infty]$.
Therefore, 
\begin{align}
    \widetilde{\mathcal{D}}_\Sigma
    =\frac{\bar\Sigma^K_\infty}{\lambda_\mathrm{gap}}
    \leq\frac{\Sigma^U_\infty}{\lambda_\mathrm{gap}}
    = \mathcal{D}_\Sigma,
\end{align}
which completes the proof of the dissipation inequality as shown in the main text.
Note that alternatively we could directly apply the thermodynamic concentration inequality of 
the main text to the particular choice $U_\Sigma$, which yields a valid but worse bound.
Namely, inserting $\| Q_\Sigma\|_{L^\infty(\mu)}\leq \Sigma^U_\infty$
from Eq.~\eqref{eq:sup_Q_Sigma_bound}
into the constant 
$\mathcal{D}=2\sqrt{\Sigma^U_\infty\|Q\|_{L^\infty(\mu)}}/\lambda_\mathrm{gap}$
from the main text gives 
$2\Sigma^U_\infty/\lambda_\mathrm{gap}=2\mathcal{D}_\Sigma$, i.e., 
an additional factor of two.
The improvement is possible since for the particular choice of $U_\Sigma$ both
$W_\Sigma$ and $\langle W_\Sigma\rangle_\mu$ are known explicitly,
and we can bound $\| W_\Sigma - \langle W_\Sigma\rangle_\mu\|_{L^\infty(\mu)}$
directly, instead of separately.

\subsection{Additional dissipation concentration inequality based on Approach 2}
We can also use the second approach to bound $\Lambda^J(k)$
to prove an additional dissipation concentration inequality.
First, recall from Eqs.~\eqref{eq:dissi_parameters1} 
and~\eqref{eq:dissi_parameters}
that for $U_\Sigma=D^{-1}\vs K^h_z$
we have $W_\Sigma=\Sigma_\mathrm{loc}K^h_z$ and
$Q_\Sigma=\Sigma_\mathrm{loc}(K^h_z)^{2}$ with
$\langle W_\Sigma\rangle_\mu=\Sigma^K$, and let us further
define
$\Sigma^{K^{2}}\equiv\langle Q_\Sigma\rangle_\mu
=\langle\Sigma_\mathrm{loc}(K^h_z)^{2}\rangle_\mu\leq\Sigma^K\leq\Sigma^U$.
Next, we establish upper bounds on the respective sup-norms via 
the same ideas from above,
\begin{align}
    \|\bar W_\Sigma\|_{L^\infty(\mu)}
    &\leq\max\bigl\{\Sigma^K,\Sigma^U_\infty-\Sigma^K\bigr\}
    \leq\Sigma^U_\infty,
    \\
    \|\bar Q_\Sigma\|_{L^\infty(\mu)}
    &\leq\max\bigl\{\Sigma^{K^2},\Sigma^U_\infty-\Sigma^{K^{2}}\bigr\}
    \leq\Sigma^U_\infty.
    \label{eq:diss_sup_bounds2}
\end{align}
The two variance bounds again follow analogously 
by pulling out one $\Sigmaloc$ factor
via its sup-norm of the probed region and are thus given by
\begin{align}
    \Var_\mu(W_\Sigma)&\leq\Sigma^U_\infty\Sigma^{K}-(\Sigma^{K})^{2},
    \label{eq:diss_var_bounds2_W}
    \\
    \Var_\mu(Q_\Sigma)&\leq\Sigma^U_\infty\Sigma^{K^{2}}-(\Sigma^{K^{2}})^{2}.
    \label{eq:diss_var_bounds2_Q}
\end{align}
Using the bounds of Eq.~\eqref{eq:diss_sup_bounds2}
for the definition of $k^\ast_{\max}$ 
allows us to identify
$k^\ast_\Sigma$ as the root of the new
simplified equation $k\Sigma^U_\infty(1+k)=\lambda_\mathrm{gap}$, 
given by
\begin{align}
    k^\ast_\Sigma
    =\frac{1}{2}\left(\sqrt{1+\frac{4\lambda_\mathrm{gap}}{\Sigma_\infty^U}}-1\right)
    \leq 
    k^\ast_{\max},
    \qquad
    \mathcal{D}_{\Sigma,2}\equiv \frac{1}{k^\ast_\Sigma}.
    \label{eq:D_Sigma_2}
\end{align}
Restricting the admissible range to $0\leq k<k^\ast_\Sigma$
and carrying out the optimization again yields the typical sub-gamma
structure and identifies
\begin{align}
    \widetilde{\mathcal{C}}_{\Sigma,2}
    \equiv 2\langle Q_\Sigma\rangle_\mu
    + \frac{2}{\lambda_\mathrm{gap}}
      \left(\sqrt{\Var_\mu(W_\Sigma)}
      + k^\ast_\Sigma\sqrt{\Var_\mu(Q_\Sigma)}\right)^{2},
\end{align}
which using the variance bounds of Eqs.~\eqref{eq:diss_var_bounds2_W} and~\eqref{eq:diss_var_bounds2_Q}
subsequently gives
\begin{align}
    \widetilde{\mathcal{C}}_{\Sigma,2}
    &\leq
    2\Sigma^{K^{2}}
    +\frac{2}{\lambda_\mathrm{gap}}
    \left(\sqrt{\Sigma^U_\infty\Sigma^{K}-(\Sigma^{K})^{2}}
    +k^\ast_\Sigma\sqrt{\Sigma^U_\infty\Sigma^{K^{2}}-(\Sigma^{K^{2}})^{2}}
    \right)^{2}
    \nonumber
    \\
    &\leq
    2\Sigma^{K}
    +\frac{2}{\lambda_\mathrm{gap}}
    \left(\sqrt{\Sigma^U_\infty\Sigma^{K}-(\Sigma^{K})^{2}}
    +k^\ast_\Sigma\sqrt{\Sigma^U_\infty\Sigma^{K}}
    \right)^{2}
    \nonumber
    \\
    &\leq 
    2\Sigma^{U}
    +\frac{2}{\lambda_\mathrm{gap}}
    \left(\sqrt{\Sigma^U_\infty\Sigma^{U}}
    +k^\ast_\Sigma\sqrt{\Sigma^U_\infty\Sigma^{U}}
    \right)^{2} 
    \nonumber
    \\
    &=
    2\Sigma^U+\frac{2}{\lambda_\mathrm{gap}}\Sigma^U_\infty\Sigma^U(1+k^\ast_\Sigma)^2
    \equiv \mathcal{C}_{\Sigma, 2},
    \label{eq:C_Sigma_2}
\end{align}
where in the second inequality we use $\Sigma^{K^2}\leq\Sigma^K$
and drop the $(\Sigma^{K^{2}})^{2}\geq 0$ term, 
and in the third inequality we use  $\Sigma^{K}\leq\Sigma^U$
and drop the $(\Sigma^K)^{2}\geq 0$ term, respectively.
Taken together, the dissipation concentration inequality~(11)
of the main text thus also holds with $\mathcal{C}_{\Sigma, 2}$
and $\mathcal{D}_{\Sigma,2}$ from Eqs.~\eqref{eq:C_Sigma_2} and~\eqref{eq:D_Sigma_2}, respectively.

\section{Proof of the inverse thermodynamic uncertainty relation}
As we outline in the main text, beyond
the tail probabilities our concentration-of-measure viewpoint
also allows us to bound the steady-state variance
of any generalized current from above.
In particular, in \cite{Arxiv_Stratonovich} we prove that, for $t>0$,
\begin{align}
  \mathrm{Var}_\mu(\overline J_t)
    \leq \frac{1}{t}\left[2\langle Q\rangle_\mu
      +\frac{2\mathrm{Var}_\mu(W)}{\lambda_\mathrm{gap}}\right]
      \leq \frac{1}{t}\left[2\|Q\|_{L^\infty(\mu)}
      +\frac{2\mathrm{Var}_\mu(W)}{\lambda_\mathrm{gap}}\right].
  \label{eq:var_current}
\end{align}
The current variance is thus bounded by the same 
local quantities $W$ and $Q$ that appear in the concentration bound.
The inverse thermodynamic uncertainty relation (iTUR)
follows directly by inserting
our previous bound on $\Var_\mu(W)$
into Eq.~\eqref{eq:var_current}, i.e., we use
$\Var_\mu(W)\leq\langle\Sigma^UQ\rangle^{\infty}_{\wedge\vee}
-\E^\mu[\overline{J}_t]^{2}$, 
which yields the refined iTUR,
\begin{align}
  \Var_\mu(\overline{J}_t)
  \leq
  \frac{2}{t}\left(\langle Q\rangle_\mu
  +\frac{\langle\Sigma^UQ\rangle^{\infty}_{\wedge\vee}
  -\E^\mu[\overline{J}_t]^{2}}{\lambda_\mathrm{gap}}\right)
  \leq
   \frac{2}{t}\left(\|Q\|_{L^\infty(\mu)}
  +\frac{\langle\Sigma^UQ\rangle^{\infty}_{\wedge\vee}
  -\E^\mu[\overline{J}_t]^{2}}{\lambda_\mathrm{gap}}\right)
  \leq
  \frac{2\|Q\|_{L^\infty(\mu)}}{t}
  \left(1+\frac{\Sigma^U}{\lambda_\mathrm{gap}}\right).
\end{align}
The first inequality yields the refined iTUR of
the main text, the second inequality follows from 
$\langle Q\rangle_\mu\leq \|Q\|_{L^\infty(\mu)}$,
and the third inequality is obtained 
by using  $\langle\Sigma^UQ\rangle^{\infty}_{\wedge\vee}
\leq\Sigma^U\|Q\|_{L^\infty(\mu)}$
and dropping the $\E^\mu[\overline{J}_t]^{2}\geq0$ term.
The last two expressions thus give the additional iTURs from App.~E.

\section{Proof of the uncertainty bounds}
Here we derive the uncertainty bounds of the main text, 
in particular, 
the confidence intervals, the crossover time, and the minimal observation time
and sample size.
Throughout, we will use the following two-sided bound for the sample mean
\begin{align}
  \mathbb{P}^\nu\bigl(|\widehat{J}_{n,t}-\E^\mu[\overline{J}_t]|\geq a\bigr)
  \leq 2N_\nu^{n}\exp\bigl[-nt\mathcal{I}^J(a)\bigr],
  \qquad
  \mathcal{I}^J(a)=\frac{\mathcal{C}}{\mathcal{D}^{2}}
  h\left(\frac{\mathcal{D}a}{\mathcal{C}}\right).
  \label{eq:uq_twosided}
\end{align}

\subsection{Confidence radius $r(n,t,\alpha)$}
The confidence radius of the main text follows by setting
the right-hand side of Eq.~\eqref{eq:uq_twosided}
equal to $\alpha\in(0,1)$ and solving for $a$.
Using $h^{-1}(v)=v+\sqrt{2v}$ as the inverse of $h(u)=1+u-\sqrt{1+2u}$
with $v\equiv\mathcal{D}^{2}\log(2N^n_\nu/\alpha)/(nt\mathcal{C})$
allows us to identify 
the confidence radius as 
\begin{align}
    r(n,t,\alpha)
    =\sqrt{\frac{2\mathcal{C}}{nt}\log\bigl(2N_\nu^{n}/\alpha\bigr)}
    + \frac{\mathcal{D}}{nt}\log\bigl(2N_\nu^{n}/\alpha\bigr),
    \label{eq:conf_radius_SM}
\end{align}
which states that
$\E^\mu[\overline{J}_t]\in[\widehat{J}_{n,t}-r,\widehat{J}_{n,t}+r]$
with probability of at least $1-\alpha$ under $\mathbb{P}^\nu$.

\subsection{Crossover time $t_c(n)$}
The two contributions of Eq.~\eqref{eq:conf_radius_SM}
have different scaling in $n$ and, as discussed in the main text, correspond
to a Gaussian and non-Gaussian regime for the error bounds, respectively.
The time at which the crossover between the regimes
occurs, $t_c$, is found when the
two terms in Eq.~\eqref{eq:conf_radius_SM} coincide.
Equating them and solving for $t$ yields
\begin{align}
    t_c(n,\alpha)
    =\frac{\mathcal{D}^{2}}{2\mathcal{C}}\frac{\log(2N^n_\nu/\alpha)}{n}
    =\frac{\Sigma^U_\infty}{\lambda_\mathrm{gap}
    (\lambda_\mathrm{gap}+\Sigma^U)}\frac{\log(2N^n_\nu/\alpha)}{n},
\end{align}
where the second equality follows from
$\mathcal{C}=2\|Q\|_{L^\infty(\mu)}(1+\Sigma^U/\lambda_\mathrm{gap})$ and
$\mathcal{D}=2\sqrt{\Sigma^U_\infty\|Q\|_{L^\infty(\mu)}}/\lambda_\mathrm{gap}$.
For $t<t_c$ the non-Gaussian
contribution $\propto(nt)^{-1}$ dominates the error bars, whereas for $t>t_c$
the Gaussian term $\propto(nt)^{-1/2}$ does.
At detailed balance $\mathcal{D}^{\mathrm{eq}}=0$, i.e., $t_c=0$
and error bars are Gaussian at all times.

\subsection{Minimal observation time $\tmin$ and sample size $\nmin$}
Fixing $n$ and a tolerance $\varepsilon>0$, the
requirement $r\leq \varepsilon$ and solving for $t$ 
yields $t\geq \tmin$ where
\begin{align}
    \tmin(n,\alpha,\varepsilon)
    =\frac{\Theta(\varepsilon)}{n}\log\bigl(2N^n_\nu/\alpha\bigr),
    \qquad
    \Theta(\varepsilon)
    \equiv\frac{\bigl(\sqrt{\mathcal{C}}
    +\sqrt{\mathcal{C}+2\mathcal{D}\varepsilon}\bigr)^{2}}{2\varepsilon^{2}}.
    \label{eq:tmin_SM}
\end{align}
Analogously, using $\log(2N^n_\nu/\alpha)=\log(2/\alpha)+n\log N_\nu$,
with requiring the same condition $r\leq \varepsilon$
but now solving for $n$ at a fixed $t$ gives
$n\geq \nmin$ with
\begin{align}
    \nmin(t,\alpha,\varepsilon)
    =\left\lceil\frac{\Theta(\varepsilon)\log(2/\alpha)}
    {t-\Theta(\varepsilon)\log N_\nu}\right\rceil,
\end{align}
which further requires that $t>\Theta(\varepsilon)\log N_\nu$, i.e., 
below this time threshold no sample size is large enough.

\subsection{Lower bounds on the minimal observation time and sample size}
Since it holds that 
$\sqrt{\mathcal{C}}\leq\sqrt{\mathcal{C}+2\mathcal{D}\varepsilon}$ and
$\sqrt{\mathcal{C}(\mathcal{C}+2\mathcal{D}\varepsilon)}\leq\mathcal{C}+\mathcal{D}\varepsilon$
the introduced $\Theta(\varepsilon)$
from Eq.~\eqref{eq:tmin_SM} can be bounded according to 
\begin{align}
    \frac{2\mathcal{C}}{\varepsilon^{2}}
    \leq\Theta(\varepsilon)
    \leq\frac{2\mathcal{C}}{\varepsilon^{2}}+\frac{2\mathcal{D}}{\varepsilon}.
    \label{eq:Theta_bound_SM}
\end{align}
Writing $\mathcal{C}=\mathcal{C}^{\mathrm{eq}}(1+\Sigma^U/\lambda_\mathrm{gap})$, where
$\mathcal{C}^{\mathrm{eq}}=2\|Q\|_{L^\infty(\mu)}$,
together with the lower bound of Eq.~\eqref{eq:Theta_bound_SM}
therefore yields the bound
\begin{align}
    \tmin(n,\alpha,\varepsilon)
    &\geq\left(1+\frac{\Sigma^U}{\lambda_\mathrm{gap}}\right)
    \frac{\Theta^{\mathrm{eq}}(\varepsilon)}{n}\log\bigl(2N^n_\nu/\alpha\bigr)
    =\left(1+\frac{\Sigma^U}{\lambda_\mathrm{gap}}\right)\tmin^{\mathrm{eq}},
    \nonumber\\
    \nmin(t,\alpha,\varepsilon)
    &\geq\left\lceil\frac{(1+\Sigma^U/\lambda_\mathrm{gap})
    \Theta^{\mathrm{eq}}(\varepsilon)\log(2/\alpha)}
    {t-(1+\Sigma^U/\lambda_\mathrm{gap})
    \Theta^{\mathrm{eq}}(\varepsilon)\log N_\nu}\right\rceil.
\end{align}
Notably, the minimal $\tmin$ and $\nmin$
required for a system out of equilibrium 
therefore exceed their equilibrium counterparts at least by a factor
determined by the observed dissipation of the underlying dynamics
$(1+\Sigma^U/\lambda_\mathrm{gap})$.

\end{document}